\documentclass[preprint,11pt]{article}
\usepackage{fullpage}
\usepackage{array,xspace,hhline,tikz,colortbl,tabularx,fixltx2e,amsmath,amsfonts,amsthm,amssymb}
\usepackage{adjustbox}
\usepackage[ruled]{algorithm2e} 
\usepackage{enumitem}
\usepackage{wrapfig}
\usepackage{bm}

\usepackage{arydshln}
\usepackage{bm}
\usepackage{blkarray}
\usepackage{booktabs}
\usepackage{caption}
\usepackage{float}
\usepackage{framed}
\usepackage{ltablex}
\usepackage{ltxtable}
\usepackage{longtable}
\usepackage{microtype}
\usepackage{mleftright}
\usepackage{multirow}
\usepackage{natbib}
\usepackage{nicefrac}
\usepackage{ifthen}
\usepackage{pifont}
\usepackage[normalem]{ulem}
\usepackage{pgfplots}
\usepackage{pgflibraryshapes}
\usepackage{verbatim,ifthen}
\usepackage{varioref}
\usepackage{xr}
\usepackage{amsthm}
\newtheorem{proposition}{Proposition}
\newtheorem{corollary}{Corollary}
\usepackage{amsmath}
\usepackage{subcaption}
\usetikzlibrary{arrows}
\usepackage{eqparbox}
\usepackage{enumitem}
\usepackage{hyperref}
\usepackage{xcolor,tcolorbox}

\definecolor{light-gray}{gray}{0.9}

\newtheorem{definition}{Definition}

\newtheorem{fact}{Fact}   

\newtheorem{lemma}{Lemma}
\newtheorem{theorem}{Theorem}
\usepackage{bm}

\definecolor{myblue}{HTML}{6397C7}
 
\newtcolorbox{resultbox}[2][]{
    halign title=flush left,
    left*=0pt, right*=0pt,
    boxsep=2pt, left=5pt, right=5pt,
    skin first=enhanced,
    skin middle=enhanced,
    skin last=enhanced,
    colframe = myblue!100,
    colback  = myblue!10,
    fonttitle=\fontfamily{ppl}\selectfont\bfseries, 
    title={\strut #2},
    #1
}

\title{The Complexity of Tullock Contests}
\author{Yu He \footnote{Northwestern Univeristy, \texttt{yuhe2030@u.northwestern.edu}. Work done while visiting the University of Chicago.}  
\and 
        Fan Yao \footnote{The University of North Carolina at Chapel Hill, \texttt{fanyao@unc.edu}. Work done while visiting the University of Chicago.}
        \and 
        Yang Yu \footnote{Massachusetts Institute of Technology, \texttt{yy5bm@virginia.edu}}
        \and 
        Xiaoyun Qiu \footnote{Dartmouth College, \texttt{xiaoyun.qiu@Dartmouth.edu}}        
         \and 
        Minming Li \footnote{City University of Hong Kong, \texttt{minming.li@cityu.edu.hk}}     
        \and  
        Haifeng Xu \footnote{University of Chicago, \texttt{haifengxu@uchicago.edu}}
 }
\date{}

\begin{document}

\maketitle 

\begin{abstract}
Despite the extensive literature on Tullock contests, computational results for the general model with heterogeneous contestants remain scarce. This paper studies the algorithmic complexity of computing a pure Nash Equilibrium (PNE) in such general Tullock contests. We find that the elasticity parameters $\{r_i\}$, which govern the returns to scale of contestants' production functions, play a decisive role in the problem's complexity. Our core conceptual insight is that the computational hardness is determined specifically by the number of contestants with \textit{medium elasticity} ($r_i \in (1, 2]$). This is illustrated by a complete set of algorithmic results under two parameter regimes:
\begin{itemize}[leftmargin=10pt]
\vspace{1mm}
    \item \textbf{Efficient Regime:} When the number of contestants with medium elasticity is logarithmically bounded by the total number of contestants ($O(\log n)$), we provide an algorithm that determines the existence of a PNE and computes an $\varepsilon$-PNE in polynomial time in both $n$ and $\log(1/\varepsilon)$ (i.e., Poly$(n,\log(1/\varepsilon))$) whenever it exists. This result generalizes classical findings for concave ($r_i \le 1$) and convex ($r_i > 2$) cases, establishing computational tractability for a broader class of mixed-elasticity contests.
\vspace{2mm}
    \item \textbf{Hard Regime:} In contrast, we show when the number of medium elasticity contestants exceed $\Omega(\log n)$, determining the existence of PNEs is NP-complete and it is impossible for any algorithm to compute an $\varepsilon$-PNE within running time Poly$(n,\log(1/\varepsilon))$. We then design a Fully Polynomial-Time Approximation Scheme (FPTAS) that computes an $\varepsilon$-PNE in Poly$(n,1/\varepsilon)$, guaranteeing efficient approximations for hard instances.
\end{itemize}
\end{abstract}

\section{Introduction}

\noindent
Tullock contest, first introduced by \citet{Tullock1980}, is a widely used  contest model where contestants expend \emph{costly} effort or resources to increase their chances of winning a reward. 
The prize structure, together with costly effort, captures many real-world applications. For instance, in pharmaceutical R\&D races, companies invest billions of dollars in developing new drugs, but only the first to invent wins the exclusive rights to sell the drug \citep{bryan2022r}. Similarly, in electoral competitions, political parties spend substantial campaign funds to influence voters, yet only one party wins the election in the end.

In addition to these widely studied examples, the advancement of cryptographic technology and digital currencies has given rise to new phenomena that can be effectively modeled and analyzed using the Tullock contest framework.  An important example is proof-of-work blockchain, a technology used to record and confirm cryptocurrency trades. In this proof-of-work system, miners (contestants) invest computational power to solve a cryptographic puzzle, the first miner to solve the puzzle is rewarded a fixed amount of cryptocurrency plus transaction fees \citep{Cong2018, Thum2018}.\footnote{A Bitcoin miner winner receives 6.25 Bitcoin (as of December 2024) plus transaction fees.}
By 2024, blockchain's global market value has exceeded \$32 billion and is expected to reach \$162.84 billion by 2027 \citep{DemandSage2024}, driven by both the high volume of cryptocurrency transactions and the entry of new types of cryptocurrencies.\footnote{For example, bitcoins alone are traded 400,000 times per day. Bitcoin, Dogecoin, and Ethereum were launched in 2009, 2013 and 2015. See \url{https://finance.yahoo.com/markets/crypto/all/} for more details.} 
Despite of this, the economic incentives of blockchain and the associated inefficiencies are not yet well understood.
For example, due to competition for computational power among miners, it is estimated that a single cryptocurrency transaction consumes approximately 80,000 times more energy than a credit card transaction \citep{IEEEBlockchain2019}.
Hence, understanding the strategic behaviors of miners is an important first step to understanding the performance of this blockchain market and identifying the potential source(s) of inefficiency. 

To this end, we study these new phenomena using a Tullock contest model with \emph{heterogeneous} contestants.
However, characterizing the equilibrium of Tullock contests in the context of blockchains presents several new challenges. In general, PNEs of games with discontinuous and non-monotonic best response is very difficult to characterize \citep{reny1999existence,olszewski2023equilibrium}. This challenge is further compounded by the fact that, as shown by \citet{ewerhart2024tullock}, Tullock equilibria may lack rational or closed-form representations even under favorable conditions, making exact symbolic computation infeasible. This analytical intractability motivates the need for computational approaches, especially in settings like blockchain where large-scale heterogeneity is the norm. In particular, the tractability of Tullock contest model in previous studies often rely on the assumptions of a small number of contestants (typically two or a few) and that they are identical \citep{skaperdas1996contest, Cornes2005}. However, these assumptions are improper in the context of blockchain --- there are likely hundreds of thousands of contestants in the blockchain contests, and these contestants usually vary in their access to information, cost of computational power (due to different locations), and computational capacities ranging
from individuals with limited resources to large-scale mining farms \citep{Cong2018, Easley2019}.\footnote{Miners depend on timely data regarding network difficulty, block rewards, and energy costs to make optimal decisions. Differences in information access can lead to significant disparities in strategic effectiveness.}
Because of this, previous results under small participation and homogeneous agents do not directly apply here. Moreover, the equilibrium analysis of Tullock contests is notoriously complicated when allowing for large-scale participation and heterogeneous players.\footnote{\cite{van2013theory} provides a unified framework to discuss the class of Tullock contests.
Using their language, while \cite{szidarovszky1997existence} prove the existence for a class of Tullock contests with decreasing return to scale, we know little about cases with increasing return to scale and heterogeneous players.
If players are identical and the total number is small, the existence and uniqueness of Nash equilibrium can be ensured; when there are many players, typically there is no pure strategy equilibrium.
If players are heterogeneous, we know very little about equilibrium existence.}

Therefore, in this paper, we explore a new direction of using a computational approach to solve general Tullock contests. We first develop a Tullock contest model that allows for an arbitrarily large number of contestants who are heterogeneous in their production capabilities and effort elasticities.Second, we show that the number of contestants $m$ whose effort elasticities fall within a specific medium regime $(1, 2)$ is the key determinant of computational complexity. Specifically, we identify a sharp phase transition: (1) Easy Case: When $m = O(\log n)$, we show that the existence of a PNE can be efficiently determined, and an $\varepsilon$-PNE can be computed in time polynomial in $n$ and $\log(1/\varepsilon)$. (2) Hard Case: When $m$ is beyond the logarithmic scale (i.e, $m = \omega(\log n)$), the problem becomes NP-complete. In this intractable regime, we prove that achieving a runtime with $\log(1/\varepsilon)$ dependence is impossible unless P = NP.

Having established these boundaries, we develop customized algorithms for both cases. For the Easy Case, we design algorithms with a runtime complexity of $\text{poly}(n, \log(1/\varepsilon))$.  For the Hard Case, we develop a Fully Polynomial-Time Approximation Scheme (FPTAS) that computes an $\varepsilon$-PNE in time polynomial in both $n$ and $1/\varepsilon$. Finally, we implement these algorithms and provide extensive numerical evaluations. Our results demonstrate that the empirical runtime behavior perfectly aligns with our $O(\frac{n^4}{\varepsilon^2}\log^2 n)$ theoretical complexity analysis, confirming the transition from tractable to intractable instances. To support further research and practical applications, we have compiled our algorithms into a Python module which is publicly available on GitHub.\footnote{To adhere to the double-blind review policy, the source code is included in the supplementary material.}

To the best of our knowledge, this is the first work to systematically study the computational complexity of Tullock contests and to provide a general and flexible toolkit for computing their PNEs. This work makes a non-trivial progress to the analysis of Tullock contests, a field that remains underdeveloped due to its inherent analytical intractability. Our toolkit can be used to study strategic behaviors in multi-agent settings and shed light on the economic implications such as allocation efficiency, centralization tendencies and sustainability of various competing settings.

\paragraph{Related Literature.}

Tullock contests have been extensively studied in economic theory as a fundamental framework for analyzing competitive resource allocation. Originally introduced to model rent-seeking behavior in political economy \citep{Tullock1980}, its theoretical robustness is underscored by \citet{baye2003strategic}, who identify conditions under which various innovation tournaments and patent-race games are strategically equivalent to the Tullock contest. Furthermore, \citet{baye1993rigging} extend this analysis to the lobbying process, modeling it as an application of the all-pay auction and highlighting the structural connections between different forms of competitive resource allocation.

Early theoretical works focused on analytical solutions under restrictive assumptions, such as a small number of contestants (often two) and symmetric players \citep{skaperdas1996contest, Cornes2005}. These assumptions enable closed-form equilibrium characterizations, particularly in settings where the effort elasticity (hereafter, ``$r$'') is $r = 1$ or $r > 2$. Recent extensions have relaxed these assumptions by incorporating asymmetry. \citet{clark1998contest} introduced heterogeneity in contestants’ production technologies, while \citet{Cornes2005} examined settings with asymmetric elasticities, demonstrating how heterogeneity affects the existence and uniqueness of equilibria. However, \citet{ewerhart2024tullock} formally shows that in general Tullock contests, closed-form solutions do not exist, even under smooth and symmetric setups. This analytical intractability motivates the need for computational approaches.

Beyond classical economic theory, the Tullock model has found renewed relevance in the digital economy. Notably, the lottery contest ($r=1$) has been established as the canonical model for Bitcoin mining and Proof-of-Work mechanisms \citep{leshno2020bitcoin, chen2019axiomatic}. Research in this domain has utilized the contest framework to analyze market structures, such as the natural tendency toward oligopoly in mining pools \citep{arnosti2022bitcoin}. Similarly, in the realm of crowdsourcing, generalized Tullock contests have been employed to model winner-take-all competitions with stochastic production \citep{cavallo2013winner} and to design competitive mechanisms that optimize cost-effectiveness \citep{rokicki2014competitive}.

From an algorithmic perspective, recent studies have begun to address the computational aspects of these contests. \citet{Elkind2024} show that under small elasticity ($r\leq 1$), continuous-time best-response dynamics globally converge to Nash equilibria. \citet{ghosh2023best} analyze best-response dynamics in Tullock contests with convex cost functions, showing that for homogeneous agents the dynamics rapidly converge to $\varepsilon$-PNE. Furthermore, \citet{ghosh2023} analyze convergence properties in lottery contests, and \citet{deng2024competition} extend the scope to competition among pairwise lottery contests. However, these algorithmic works largely focus on specific elasticity regimes (often $r=1$) or specific dynamic settings. Our work complements this literature by addressing the computational complexity of equilibrium in general heterogeneous Tullock contests without imposing restrictions on elasticity or cost convexity.

Finally, while methods have been widely applied in other domains, particularly in single-agent decision-making or non-competitive settings, such as optimal resource allocation and learning-based optimization \citep{Mart1990, even2009convergence}. However, these approaches do not directly address the complexities introduced by strategic interactions among multiple agents. \citet{ghosh2023} explore best-response dynamics in contests, providing tight bounds on convergence rates across elasticity regimes, but they focus on homogeneous players. We extend the computational techniques to Tullock contest framework, incorporating heterogeneous elasticity and asymmetry to analyze equilibria in decentralized and competitive environments. Our findings contribute to the existing literature on computational complexity in multi-agent competing settings, underscoring the role of elasticity in determining equilibrium properties.

Applications of Tullock contests extend to various domains, including blockchain and decentralized finance systems. Blockchain mining, as modeled in proof-of-work mechanisms, exemplifies a highly competitive environment involving thousands of heterogeneous participants with varying computational resources and costs \citep{Cong2018, Easley2019}. These settings pose unique challenges, such as resource centralization and inefficiency, which are not addressed by traditional contest models. Our work bridges this gap by providing a computational framework for analyzing equilibrium outcomes in blockchain-inspired contests, contributing to the design and optimization of decentralized systems. Additionally, insights from our study can inform broader applications in governance, resource allocation, and decentralized competitive systems \citep{fu2016disclosure}.

\section{The Tullock Contest and Overview of Our Results}

This paper studies the complexity of \emph{Tullock contests}, a widely adopted contest model due to the seminal work of \citet{Tullock1980}. A Tullock contest consists of $ n $ contestants ($ n \geq 2 $) competing for a reward of amount $ R (>0)$. The action of each contestant $ i \in [n] $ is to pick a non-negative \emph{effort level} $ x_i  \in \mathbb{R}_{\geq 0} $. Each contestant $ i $ is characterized by a \emph{production function} $ f_i(x_i) := a_i x_i^{r_i}$,   specified by two positive parameters $ (a_i, r_i) \in \mathbb{R}^2_{>0}$; $ a_i$ describes $ i $’s \emph{production efficiency}, and $ r_i  $ captures the \emph{elasticity} of effort \citep{Tullock1980}. Notably, many previous works restrict $r_i$ to be the same for all contestants. Here we relax this restriction and allow heterogeneous elasticities. As we will show later, this generalization would not fundamentally  change the computational complexity of the contest. 

In Tullock contests, all contestants move simultaneously. Contestant $i$'s winning probability is proportional to their own production $f_i(x_i)$, defined as $p_i = \frac{f_i(x_i)}{\sum_{j=1}^{n} f_j(x_j)} $ where $f_i(x_i) = a_i x_i^{r_i}$.\footnote{This is well-defined, except for the extreme case with $x_j=0$ for all $j$. In this case, we assume $p_j = 1/n$ for all $j$. } Hence, the payoff to each contestant $i$, assuming linear cost in effort, is:

\begin{equation}\label{eq:tullock-utility}
u_i(x_i, \boldsymbol{x}_{-i}) = \frac{ a_i x_i^{r_i}}{\sum_{j=1}^n a_j x_j^{r_j}} \cdot R - x_i 
\end{equation}

A Tullock contest game is thus fully specified by $ 2n+1 $ positive numbers,  $ \mathcal{G} = \{ (a_i, r_i) \}_{i=1}^n \cup \{ R \} $, which we refer to as an instance $\mathcal{G}$. Given any $\mathcal{G}$, we study the complexity of computing an $\varepsilon$-PNE of $\mathcal{G}$ (as defined in section~\ref{refulation}) or assert that no pure Nash equilibrium (as defined below) exists. Throughout our complexity analysis, we adopt the standard assumption that the input parameters $a_i$ and $r_i$ are strictly bounded. Specifically, we assume there exist universal positive bounds independent of $n$ (say, $[\underline{c}, \bar{c}]$ where $0 < \underline{c} \le \bar{c} < \infty$) that restrict the valid range of all $a_i$ and $r_i$. This assumption is justified on two grounds: first, $a_i$ and $r_i$ represent the intrinsic capability of a contestant and the structural technology of the contest, respectively, which are local properties whose feasible magnitudes do not naturally scale with the population size $n$; second, bounding these parameters away from zero and infinity ensures that all contestants are non-trivial participants and that the objective functions remain well-behaved, thereby preventing numerical instability.

Like most previous works about Tullock contests \citep{Cornes2005,Elkind2024,ewerhart2024tullock,ghosh2023best}, we also focus on Pure Nash Equilibrium (PNE) in this paper, as naturally defined below.  
\begin{definition}[Pure Nash equilibrium]\label{def:eq} An action profile $\boldsymbol{x}^* = (x_1^*,x_2^*,\cdots,x_n^*)$ is a Pure-strategy Nash Equilibrium (PNE) if and only if it satisfies
\[
u_i(x_i^*,\textbf{x}^*_{-i}) \geq u_i(x_i,\textbf{x}^*_{-i}) \qquad \forall x_i \in \mathbb{R}_{\geq 0}, \, \, \forall i \in [n]. 
\]
\end{definition}

\paragraph{Regimes of the Elasticity.} An important conceptual insight of our complexity-theoretic study is that the regime of the parameter $r_i$ turns out to govern the computational complexity of the contest's PNE. The parameter $r_i$ is conventionally referred to as \emph{elasticity}, which measures the relative change in a contestant's payoff function with respect to the change in their effort. Formally, observe that  can be expressed as   $r_i = \frac{\Delta f_i / f_i}{\Delta x_i / x_i}$, where $\Delta f_i, \Delta x_i$ represents the infinitesimal change in the contestant's payoff function $f_i$  and  effort level $x_i$, respectively. The following three regimes of elasticity parameter turn out to be central to the game's equilibrium complexity. 

\begin{definition}[Small/Medium/Large Elasticity] 
We say $r_i$ is of (1) \textbf{small elasticity} if $r_i \in (0, 1]$; (2) \textbf{medium elasticity} if   $r_i \in (1, 2]$; (3) \textbf{large elasticity} if  $r_i \in (2, \infty)$.  
\end{definition}
Recall that $f_i(x_i) =a_i x_i^{r_i}$. Hence, the production function $f_i$ is concave under small elasticity, and convex under medium and large elasticity. 

The following useful lemma shows that, to study the complexity of PNE under different  regimes, it is without loss of generality to assume $R=1$ since for any given game $ \mathcal{G} = \{ (a_i, r_i) \}_{i=1}^n \cup \{ R \} $, its  PNE set is the same (up to a rescaling factor $R>0$) as the PNE set of another game $\mathcal{G}'$ that has the same elasticity parameter. The proof constructs $a'_i = a_i R^{r_i}, \forall i$, and is deferred to Appendix \ref{append:lem:R=1}.    
\begin{lemma}\label{lem:R=1}
For any Tullock game instance   $ \mathcal{G} = \{ (a_i, r_i) \}_{i=1}^n \cup \{ R \} $, there exists another instance $\mathcal{G}' = \{ (a'_i, r_i) \}_{i=1}^n \cup \{ 1 \}  $ such that an action profile $\boldsymbol{x}^* = (x_1^*,x_2^*,\cdots,x_n^*)$ is a PNE in $\mathcal{G}$ \emph{if and only if } $\boldsymbol{x}^*/R = (\frac{x_1^*}{R},\frac{x_2^*}{R},\cdots,\frac{x_n^*}{R})$  is a PNE in $\mathcal{G}'$. 
\end{lemma}

Thanks to Lemma \ref{lem:R=1}, we henceforth assume $R=1$ throughout, in which case we shall have $u_i \in [0, 1]$ at any equilibrium. Notably, this also gives a natural utility range when we develop additive $\varepsilon$-approximation schemes. The main results of this paper can be summarized as follows: 

\begin{resultbox}{Summary of Main Results}
 Our main result is a full characterization of the complexity of computing PNEs in Tullock contests. Let $m = |\{ i: r_i \in (1, 2] \}|$ denote the number of contestants with medium elasticity. It turns out that the computational complexity is strictly governed by $m$:
 \begin{itemize}
    \item \textbf{Efficient Regime:} When $m = O(\log n)$, the problem is computationally tractable. We provide an algorithm that determines the \textit{exact} existence of a PNE in polynomial time and, if one exists, computes an $\varepsilon$-PNE efficiently in time polynomial in $n$ and $\log(1/\varepsilon)$.
    
    \item \textbf{Hard Regime:} When $m$ is not logarithmically bounded (specifically $m = \Omega(\log n)$), determining the existence of a PNE is \textbf{NP-complete}. In addition, we prove a stronger inapproximability result: unless $P=NP$, no algorithm can compute an $\varepsilon$-PNE in time polynomial in $\log(1/\varepsilon)$ (i.e., high-precision approximation is impossible).
    
    \item \textbf{Algorithmic Resolution (FPTAS):} To address this intractability, we design a Fully Polynomial-Time Approximation Scheme (FPTAS). This algorithm circumvents the hardness barrier by computing an $\varepsilon$-PNE in time polynomial in $n$ and $1/\varepsilon$, achieving a strong possible approximation for the general regime.
 \end{itemize}
\end{resultbox}


\paragraph{Remarks on Tullock Contests.} A few remarks are worth mentioning about Tullock contests. First, there are two popular special cases: when $r_i = 1, \forall i$, it is well-known as the  \emph{lottery contest}. Here, each contestant’s winning probability is proportional to their monetary contribution. This is the initial format of Tullock contests as studied in \citep{Tullock1980} and has been shown to naturally capture winner selection probabilities in bitcoin mining \cite{leshno2020bitcoin,chen2019axiomatic,arnosti2022bitcoin}. Lottery contests' algorithmic studies  have also received much recent interest  \citep{ghosh2023best,deng2024competition}. Another popular special case is when $r_i \to \infty, \forall i$. In this case, the contestant with the largest $r_i$ wins with probability $1$. This effectively becomes a ``winner-take-all'' contest, which has found wide applications in crowdsourcing \cite{cavallo2013winner,rokicki2014competitive}.  

Second, a natural question one might have is why the particular format of winning probabilities and payoff in Equation \eqref{eq:tullock-utility} is of particular interest. This turns out to be a well-understood question. Classic works of \citet{skaperdas1996contest} and \citet{clark1998contest} show that the Equation \eqref{eq:tullock-utility} is the \emph{unique} utility form that satisfies the following three   natural axioms: (1) contestant $i$'s winning probability strictly increases in $x_i$ and decreases in $x_j$ for any $j \not = i$; (2) the choice between two alternatives is independent of an unchosen third alternative (also widely known as Luce's Choice Axiom \citep{luce1959individual} or independence from irrelevant alternatives); (3) if every contestant simultaneously scales up their effort level by any factor $\lambda>0$, each contestant's winning probability would not change. In other words, the Tullock contest is not merely a convenient modeling choice but the only contest format consistent with these basic and widely accepted principles.


\section{Warm-up for Algorithmics: Useful Properties of Equilibria  }

\label{sec:warmup}

Before analyzing the computational complexity of Tullock contests, we first establish the structural foundations of their PNEs. While closed-form characterizations of equilibria are generally intractable in heterogeneous settings, we can leverage specific structural properties of the best-response functions to design tractable solution methods. 

Our approach relies on a \emph{change of variables} strategy. Rather than working with high-dimensional effort profile $\boldsymbol{x}$, we reformulate the problem in terms of a single scalar quantity: the \emph{aggregate production} $A = \sum_{j=1}^{n} y_j$. This transformation effectively decouples strategic interactions and allows us to characterize the PNE as a consistency check on $A$.

\subsection{Equivalent Contest Re-formulation}\label{refulation}

For analytical convenience, we re-formulate the utility functions by shifting the decision variable from effort levels to production shares. Let $y_i \equiv f_i(x_i) = a_i x_i^{r_i}$ denote the production of contestant $i$, and let $A = \sum_{j} y_j$ be the aggregate production. This formulation leads to the following natural concept that is crucial to our analysis.

\begin{definition}[Action Share]
For any contestant $i$, the action share $\sigma_i$ is defined as the proportion of the aggregate production attributed to $i$: 
$$\sigma_i = \frac{y_i}{A}.$$
\end{definition}

The action share $\sigma_i$ represents the probability that contestant $i$ wins the contest. By substituting $y_i = \sigma_i A$ into the inverse production function $x_i = (y_i/a_i)^{1/r_i}$, we can express the cost of maintaining share $\sigma_i$ directly in terms of $A$. Consequently, the contestant's payoff function is re-formulated as:
\[
\pi_i(A, \sigma_i) = \sigma_i - \left(\frac{\sigma_i A}{a_i}\right)^{\frac{1}{r_i}}.
\]
For notation brevity, we sometimes denote the cost term as $g_i(\sigma_i A) = (\frac{\sigma_i A}{a_i})^{1/r_i}$. 
It is straightforward to verify that this re-formulation is equivalent to the original game since $x_i$ and the pair $(A, \sigma_i)$ are in one-to-one correspondence. We can now derive the PNE properties for the new notations $A$ and $\boldsymbol{\sigma}=(\sigma_1,\cdots,\sigma_n)$.

\begin{definition}[PNE in Terms of Aggregate Action and Action Shares] \label{def:equiv_A_sigma}
A pair $(A^*, \boldsymbol{\sigma}^*)$ constitutes a PNE if and only if for every contestant $i$, the share $\sigma_i^*$ maximizes their payoff given the fixed production of all other contestants. Formally:
\[
    \pi_i(A^*, \sigma_i^*) \geq \pi_i(A', \sigma_i'), \quad \forall \sigma_i' \in [0, 1),
\]
where the deviation aggregate production $A'$ is determined by the consistency condition:
\[
    A'(1 - \sigma_i') = A^*(1 - \sigma_i^*).
\]
\end{definition}

Definition~\ref{def:equiv_A_sigma} restates the standard notion of PNE under our change-of-variables framework. The key observation is that each contestant’s payoff depends on others only through their aggregate production, which equals $A(1 - \sigma_i)$. Thus, when contestant $i$ deviates, the consistency condition ensures that the aggregate contribution of the remaining contestants remains fixed. From this point forward, our focus shifts to determining the pair $(A^*, \boldsymbol{\sigma}^*)$, as it fully characterizes the PNE.

\subsection{Best Response Dynamics and Participation}

We first examine the behavior of a single contestant $i$ facing a fixed aggregate production $A$. The contestant chooses an action share $\sigma_i \in [0, 1]$ to maximize $\pi_i(A, \sigma_i)$.

The \textit{first-order condition (FOC)} for this maximization, $\frac{\partial \pi_i}{\partial \sigma_i} = 0$, yields a candidate solution denoted by $k(A; a_i, r_i)$. As shown in \citet{Cornes2005}, the functional form of $k$ depends on the elasticity regime:
\begin{align}\label{foc}
    k(A; a_i, r_i) \text{ solves }
    \begin{cases}
        (1 - \sigma) - \frac{1}{r_i} \left( \frac{A}{a_i} \right)^{\frac{1}{r_i}} \sigma^{\frac{1}{r_i}-1} = 0 & \text{if } i \in \mathcal{I}^1 \quad (r_i \le 1), \\
        a_i r_i^{r_i} (1 - \sigma)^{r_i} \sigma^{r_i - 1} - A = 0 & \text{if } i \in \mathcal{I}^2 \quad (r_i > 1).
    \end{cases}
\end{align}
However, satisfying the FOC is necessary but not sufficient. A valid best response must also satisfy the \textit{participation constraint}: the contestant will only exert effort if the resulting payoff is non-negative. This introduces a critical distinction between the two contestant types. We will use $b_1(A,\sigma_i;a_i,r_i)$, $b_2(A,\sigma_i;a_i,r_i)$ to represent the FOC and $k_1(A;a_i,r_i)$, $k_2(A;a_i,r_i)$ to represent the solution when $r_i \leq 1$ and $r_i > 1$.

\begin{fact}[Structure of Best Response \citep{Cornes2005}] \label{fact:br_structure}
Let $\sigma_i^{BR}(A)$ denote the best-response action share of contestant $i$ given aggregate production $A$, which has the following characterizations:
\begin{enumerate}
    \item \textbf{contestants in $\mathcal{I}^1$ ($r_i \le 1$): Continuous Participation.} \\
    These contestants have concave production functions (diminishing returns). Their best response is unique, continuous, and given by:
    \[
    \sigma_i^{BR}(A) = \max \left\{ 0, k_1(A; a_i, r_i) \right\}.
    \]
    Specifically, they participate ($\sigma_i > 0$) for all $A$ below a trivial upper bound $A_{\max}$, where $A_{\max} = \sum_i a_i$. This bound arises because contestants will not exert effort exceeding the reward ($x_i \le 1$), which implies that the maximum individual production is limited to $a_i$.
    
    \item \textbf{contestants in $\mathcal{I}^2$ ($r_i > 1$): Threshold-Based Participation.} \\
    These contestants have convex production functions. Their participation is governed by two critical thresholds, $\underline{A}_i$ and $\overline{A}_i$:
    \begin{equation}\label{eq:thresholds}
    \underline{A}_i \equiv \frac{a_i (r_i - 1)^{r_i-1}}{r_i^{r_i}}, \quad \overline{A}_i \equiv r_i \cdot \underline{A}_i.
    \end{equation}
    The best response is a \textbf{correspondence} that may take multiple values depending on the formation of $A$:
    \[
    \sigma_i^{BR}(A) \in 
    \begin{cases}
    \{k_2(A; a_i, r_i)\} & \text{if } A \le \underline{A}_i \quad \text{(Must Participate)}, \\
    \{0, k_2(A; a_i, r_i)\} & \text{if } \underline{A}_i < A \le \overline{A}_i \quad \text{(Ambiguous Region)}, \\
    \{0\} & \text{if } A > \overline{A}_i \quad \text{(Drop Out)}.
    \end{cases}
    \]
\end{enumerate}
\end{fact}

\paragraph{Key Insight: Discontinuous Drop-out and Ambiguity.}
The behavior of contestants in $\mathcal{I}^2$ (Fact \ref{fact:br_structure}.2) is the source of the algorithmic complexity addressed in this paper. Unlike contestants in $\mathcal{I}^1$ whose efforts decay smoothly, contestants with $r_i > 1$ face a discontinuous participation decision governed by whether the aggregate production includes their contribution:

\begin{itemize}
    \item \textbf{The Drop-out Effect ($A > \overline{A}_i$):} The threshold $\overline{A}_i$ represents the limit of the \textit{total} production (including contestant $i$). If the aggregate production $A$ exceeds this level, contestant $i$ cannot secure a non-negative payoff even with their optimal share, forcing them to exit ($\sigma_i = 0$).
    
    \item \textbf{The Ambiguity Region $[\underline{A}_i, \overline{A}_i]$:} The threshold $\underline{A}_i$ represents the entry limit based on the aggregate production of \textit{others} ($A_{-i} = A - \sigma_i A$). If the production of all other contestants exceeds $\underline{A}_i$, it is optimal for contestant $i$ to remain inactive. The ambiguity in this interval arises because the contestant's active status is not uniquely determined by $A$ alone; rather, it depends on the composition of $A$. Specifically, we do not know a priori whether $A$ is formed \textit{with} contestant $i$ (requiring $A \le \overline{A}_i$) or \textit{without} contestant $i$ (requiring $A_{-i} \ge \underline{A}_i$). This creates a combinatorial uncertainty regarding the set of active contestants.
\end{itemize}

Having established the best response structure, we briefly note the monotonicity properties which are crucial for our algorithmic design. To facilitate the analysis, we partition the set of contestants $\mathcal{I}$ into two subsets based on their elasticity: let $\mathcal{I}^1 = \{i \in \mathcal{I} \mid r_i \le 1\}$ denote contestants with concave (or linear) production, and $\mathcal{I}^2 = \{i \in \mathcal{I} \mid r_i > 1\}$ denote contestants with convex production.

\begin{figure}
    \centering
    \includegraphics[width=0.9\linewidth]{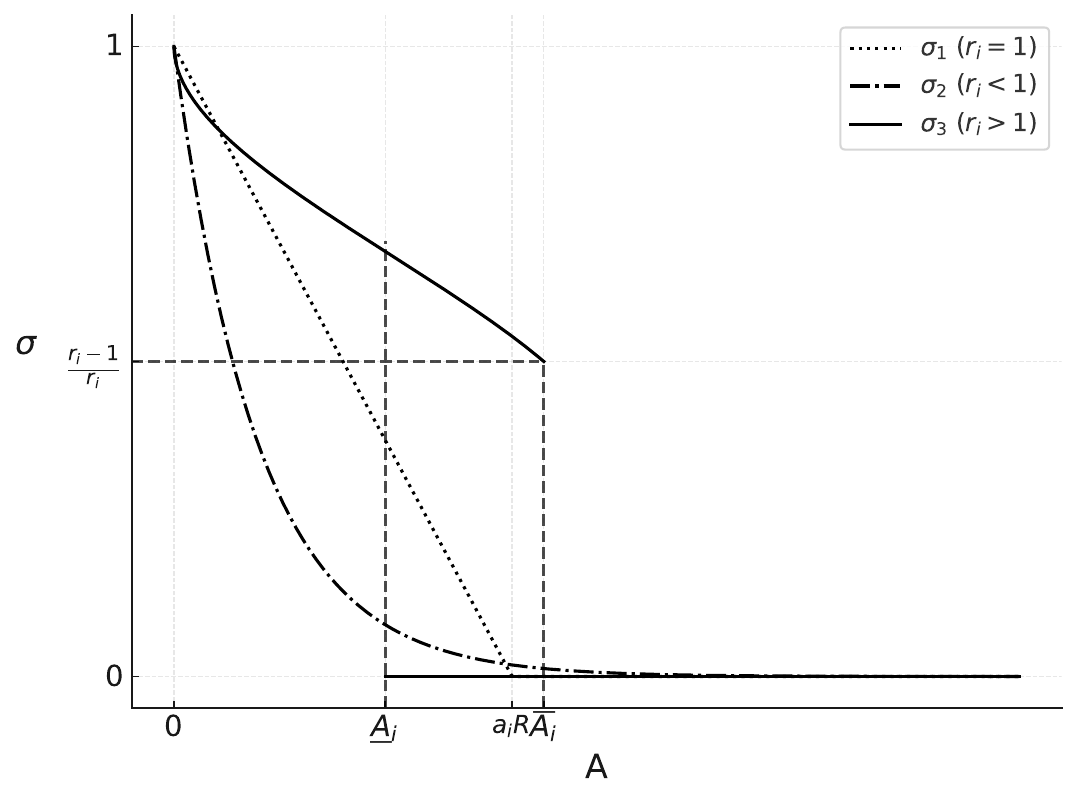}
    \caption{Best response action share $\sigma_i$ as a function of aggregate production $A$. Note the discontinuous drop to zero for $r_i > 1$ compared to the smooth decay for $r_i \le 1$. (Note that when $A = \overline{A_i}$ for $i \in \mathcal{I}^2$, $\sigma_i = \frac{r_i-1}{r_i}$)}
        \label{fig1}
\end{figure}
\begin{proposition}[Monotonicity of Best Response] \label{sa}
The first-order candidate share $k(A)$ is strictly decreasing in $A$. This implies:
\begin{enumerate}
    \item \textbf{Small Elasticity ($r_i \leq 1$):} The best response $\sigma_i^{BR}(A)$ is continuous and strictly decreasing.
    \item \textbf{Medium and Large Elasticity ($r_i > 1$):} The best response $\sigma_i^{BR}(A)$ is strictly decreasing on its support from $1$ to $\frac{r_i-i}{r_i}$(where $\sigma_i > 0$), but exhibits a discontinuous jump to $0$ within the interval $[\underline{A}_i, \overline{A}_i]$.
\end{enumerate}
\end{proposition}

\noindent \textit{Remark:}
Proposition \ref{sa} highlights a key algorithmic insight: the mathematical "engine" of the best response ($k(A)$) is well-behaved (monotonic) for all contestant types. The complexity in the medium and large elasticity regime arises solely from the discrete, combinatorial decision of \textit{whether} to participate (the jump to 0), not from the continuous decision of \textit{how much} to bid conditional on participation. Figure~\ref{fig1} visualizes these dynamics. The detailed proof is shown in Appendix ~\ref{Prof.sa}.


\subsection{General PNE Characterization}

We now assemble these components to characterize the PNE. We argue that a PNE can be fully determined by a set of \textbf{active contestants} $\mathcal{I}^A \subseteq \mathcal{I}$ (those contributing $\sigma_i^* > 0$) and a consistent aggregate production level $A^*$, as shown in the following Proposition, which unifies the PNE conditions into three structural constraints: Market Clearing, Internal Stability, and External Stability.

\begin{proposition}[PNE Conditions] \label{PNEProp} 
A triplet $\{A^*, \mathcal{I}^A, \boldsymbol{\sigma}^*\}$ constitutes a PNE if and only if the following conditions are satisfied:

\begin{enumerate}
    \item \textbf{Best Response Consistency:} 
    For every active contestant $i \in \mathcal{I}^A$, the share $\sigma_i^*$ must satisfy the first-order condition given $A^*$. Specifically: $\sigma_i^* = k(A^*; a_i, r_i)$,
    which implies that $\sigma_i^*$ solves the stationarity equations defined in Eq.~\eqref{foc}:
    \begin{align*}
        b_1(A^*, \sigma_i^*; a_i, r_i) &= 0, \quad \forall i \in \mathcal{I}^A \cap \mathcal{I}^1, \\
        b_2(A^*, \sigma_i^*; a_i, r_i) &= 0, \quad \forall i \in \mathcal{I}^A \cap \mathcal{I}^2.
    \end{align*}

    \item \textbf{Internal Stability (No Exit):} 
    Every active contestant must prefer participation to exiting. This imposes a collective upper bound on $A^*$:
    \begin{equation} \label{eq:internal}
        A^* \leq \min_{i \in \mathcal{I}^A \cap \mathcal{I}^2} \overline{A}_i.
    \end{equation}
    (Note: For $i \in \mathcal{I}^1$, participation is guaranteed provided $A^*$ allows for a valid solution to $b_1=0$).

    \item \textbf{External Stability (No Entry):} 
    Every inactive contestant must prefer staying out to entering. This imposes a collective lower bound on $A^*$:
    \begin{equation} \label{eq:external}
         A^* \geq \max_{j \in \mathcal{I}^2 \setminus \mathcal{I}^A} \underline{A}_j.
    \end{equation}

    \item \textbf{Market Clearing:} 
    The individual shares of active contestants must sum to unity:
    \begin{equation} \label{eq:clearing}
        \sum_{i \in \mathcal{I}^A} \sigma_i^* = 1 \quad \left( \iff \sum_{i \in \mathcal{I}^A} k(A^*; a_i, r_i) = 1 \right).
    \end{equation}
 \end{enumerate}
\end{proposition}

Proposition \ref{PNEProp} provides the structural foundation for our algorithmic analysis. It reveals that the problem of identifying a PNE effectively couples a continuous search for $A$ with a discrete subset selection problem. Specifically, for any hypothesized aggregate production $A$, the participation thresholds partition the contestants into three categories: those who \textit{must} be active (i.e., $A < \underline{A}_i$), those who \textit{must} be inactive (i.e., $A > \overline{A}_i$), and those in the \textbf{ambiguous region} (i.e., $\underline{A}_i \le A \le \overline{A}_i$) who may strategically choose to be either active or inactive. The detailed proof is shown in Appendix ~\ref{Prof.PNE}.

\begin{figure}
    \centering
    \includegraphics[width=0.8\linewidth]{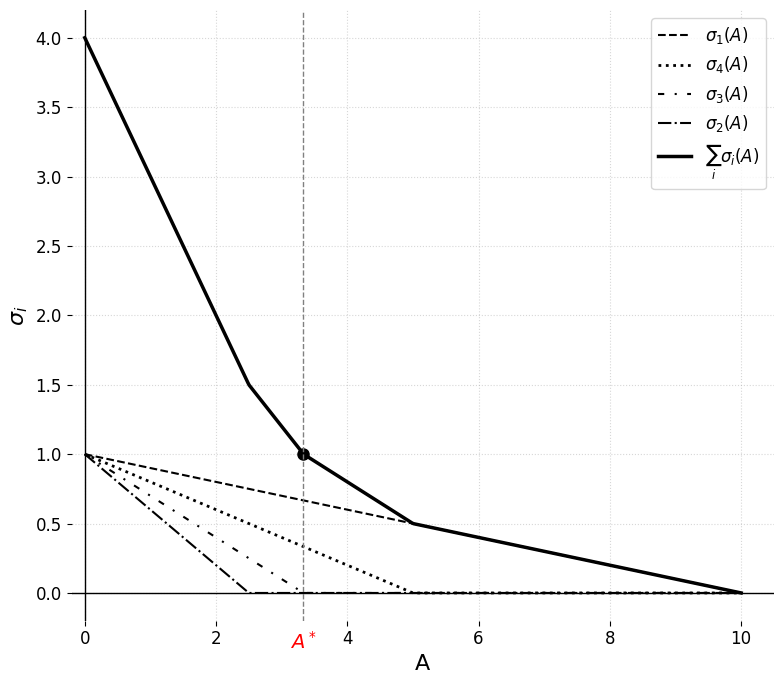}
    \caption{Best response action share as a function of aggregate production $A$ when \( r_i = 1 \). The strict monotonicity allows for efficient root-finding.}
    \label{fig2}
\end{figure}
Consequently, finding an PNE necessitates resolving this combinatorial uncertainty: the algorithm must identify a valid configuration of the ambiguous contestants such that, when combined with the mandatory participants, their cumulative optimal shares sum exactly to unity at the consistent level $A$. This interplay between the continuous state variable $A$ and the discrete participation choices constitutes the core computational challenge addressed in the subsequent sections.

\section{Efficiently Solvable Regimes}
\label{sec:efficient}

One of our key findings is that the computational complexity of finding a PNE in Tullock contests is fundamentally governed by the heterogeneity of the elasticity parameters $\{r_i\}$. While the general problem is intractable, we identify broad and practical-relevant regimes where the PNE can be computed efficiently.

Our core insight is that the hardness of the problem stems primarily from the number of contestants with \textit{medium elasticity} (i.e., $r_i \in (1, 2]$). Let $m = |\{i : r_i \in (1, 2]\}|$ denote the count of such contestants. In this section, we prove that whenever $m$ is logarithmically bounded (i.e., $m = O(\log n)$), the PNE computation is tractable.

It is important to note, however, that computational tractability does not imply the existence of a closed form solution. As formally shown by \citet{ewerhart2024tullock}, PNE in Tullock contests generally lack rational or closed-form representations. For instance, even in a simple two-contestant contest with $r_1=0.5$ and $r_2=1$, the unique PNE effort levels are irrational numbers (see Appendix~\ref{proof_easy_2} for the detailed proof). Consequently, our goal is to algorithmically compute the PNE to arbitrary precision efficiently, through finding an $\varepsilon$-pure Nash equilibrium ($\varepsilon$-PNE)--- a commonly adopted solution concept in game theory when the exact PNE is intractable \citep{daskalakis2006note}--- as shown in the following Definition \ref{def:eps_approx}. We show that an $\varepsilon$-PNE can be achieved in polynomial time in the input size $n$ and the precision parameter $\log(1/\varepsilon)$.
\begin{definition}\label{def:eps_approx}[$\varepsilon$-pure Nash equilibrium]
A strategy profile $ \mathbf{x}^* = (x_1^*, x_2^*, \dots, x_n^*) $ is an $\varepsilon$-pure Nash equilibrium ($\varepsilon$-PNE) if for every contestant $ i $, the effort level $ x_i^* $ satisfies:
\[
u_i(x_i^*, \mathbf{x}_{-i}^*) \geq \max_{x_i \geq 0} u_i(x_i, \mathbf{x}_{-i}^*) - \varepsilon.
\]
\end{definition}

\subsection{The Baseline: Efficient Approximation for Small Elasticity ($ r_i \le 1 , \forall i $)}
\label{easy_case}

We begin by analyzing the regime where all contestants possess concave production functions ($r_i \leq 1$). This setting constitutes the structural baseline of Tullock contests. Theoretically, this class falls under the umbrella of \textit{strictly monotone games} \citep{rosen1965existence} and satisfies the property of \textit{diagonal strict concavity} \citep{szidarovszky1997existence, even2009convergence}. Consequently, the existence and uniqueness of the PNE is guaranteed.

\paragraph{Algorithmic Implication: Monotonicity.}
While these existence results are classical, our algorithmic focus is on the specific structure of the aggregate production function. As established in Proposition~\ref{sa}, for $r_i \le 1$, each contestants's best-response share $\sigma_i(A)$ is strictly decreasing in $A$. This implies that the aggregate best-response share function, $\sum \sigma_i(A)$, is strictly monotonic.
Therefore, finding the unique PNE $A^*$ reduces to finding the unique root of $\sum \sigma_i(A) = 1$.

Figure~\ref{fig2} illustrates this favorable property. The strict monotonicity ensures that standard root-finding techniques, such as \emph{Binary Search}, can efficiently converge to the PNE.

Next, we develop two efficient algorithms to compute \(\varepsilon\)-PNEs under small elasticity, leveraging the strict monotonicity of best-response mappings in Tullock contests.

\textbf{Bisection Method.}  
The first algorithm applies a bisection procedure to the aggregate production level \( A \). Because the PNE condition \( \sum_{i=1}^n \sigma_i(A) = 1 \) has a unique solution and each \( \sigma_i(A) \) is strictly decreasing, bisection converges efficiently. We terminate when \( \big| \sum_i \sigma_i - 1 \big| \leq \delta(\varepsilon) \), ensuring that the resulting profile satisfies the \(\varepsilon\)-PNE condition.

\textbf{Multi-Agent Mirror Descent.}  
We also implement a multi-agent learning algorithm, Multi-Agent Mirror Descent (MAMD), for finding an $\varepsilon$-PNE. First proposed by \citep{bravo2018bandit}, MAMD is an iterative algorithm for finding equilibria in monotone games, achieved by letting players update their strategies sequentially to improve their utilities using noisy observations. In our setting, we adopt a variant of MAMD \citep{yao2024human}, by allowing contestants to leverage the utility gradient information. Specifically, each contestant updates their strategy via mirror descent with respect to the gradient of their own utility. The algorithm converges to an \(\varepsilon\)-PNE under mild regularity assumptions and is particularly scalable in high-dimensional or parallelizable settings.

Both methods achieve polynomial-time convergence in \( O(\log \frac{1}{\varepsilon}) \), with full correctness proofs deferred to the appendix ~\ref{app:approx}. Together, they demonstrate that the small elasticity regime admits efficient and practical algorithms for $\varepsilon$-PNE computation.

\subsection{Structural Properties of Large Elasticity ($ r_i > 2, \forall i $)}
In this regime, the elasticity parameters $r_i$ are greater than or equal to 2, leading to convex production functions that exhibit increasing marginal returns on effort. Unlike the concave regime, where the PNE is distributed among contestants, the convex regime typically results in  only one contestant remaining active in the PNE.

\begin{proposition}\label{None}
    A Tullock contest with $ r_i > 2 $ for all contestants admits no PNE.
\end{proposition}
To provide an intuitive explanation, consider the condition $ r_i > 2 $. For any active contestant $ i $, by proposition~\ref{sa}, their action share at a PNE $ \sigma_i $ must exceed $ \frac{r_i - 1}{r_i} > \frac{1}{2}$. This implies that if a PNE exists, at most one contestant can be active, as the total action share across all contestants must sum to 1. However, for a single active contestant, their action share would have to be $ \sigma_i = 1 $, which requires the aggregate production $ A $ to approach 0. But when $ A$ approaches $0$, all contestants are active since the cost of participation vanishes, contradicting the assumption that only one contestant remains active. This contradiction confirms that no configuration satisfies the PNE conditions, demonstrating the absence of PNEs under $ r_i > 2 $. Full details are provided in Appendix \ref{A}.

\begin{corollary}\label{act}
    Any PNE of a Tullock contest contains at most one active contestant with $ r_i > 2 $.
\end{corollary}
 
The restriction to a single active contestant follows from the requirement that their action share exceeds $\frac{1}{2}$, leaving no room for additional active contestants with large elasticity.

\subsection{A Unified Polynomial-Time Algorithm for Bounded Medium Elasticity ($ r_i \in (1,2], \forall i $)}
\label{sec:unified_algo}

Building on the structural properties established above, we now address the general setting containing contestants from all elasticity regimes. We consider the case where the number of medium elasticity contestants is bounded by $m = O(\log n)$.

The challenge in this general setting is determining the set of active contestants, denoted by $\mathcal{I}^A$. Recall that finding a PNE is equivalent to identifying a valid pair $(\mathcal{I}^A, A^*)$. Our strategy relies on a combinatorial search over the possible configurations of $\mathcal{I}^A$, which is made efficient by the specific behaviors of different elasticity types:

\begin{itemize}
    \item \textbf{Small Elasticity ($r_i \leq 1$):} These contestants are "persistent". They do not need to be enumerated because, given any candidate $A^*$, their participation status is deterministic (they are active if $A^* < A_{\max}$ and their marginal utility is positive).
    \item \textbf{Large Elasticity ($r_i > 2$):} These contestants are "exclusive". By Corollary \ref{act}, at most one contestant from this set can be active in any PNE. This limits the search space for this group to $O(n)$ possibilities.
    \item \textbf{Medium Elasticity ($r_i \in (1, 2]$):} These contestants exhibit "ambiguous" participation behavior. However, since their count $m$ is small ($O(\log n)$), we can afford to exhaustively enumerate their active subsets.
\end{itemize}

Based on this decomposition, we propose a unified algorithm (Algorithm 3) that effectively determines the existence of a PNE and computes it if one exists.

\begin{theorem}[Efficiency of the Unified Algorithm] \label{the2}
    Consider any Tullock contest instance where the number of contestants with medium elasticity ($r_i \in (1,2]$) is bounded by $m = O(\log n)$. There exists a algorithm that:
    \begin{enumerate}
        \item \textbf{Exact Existence Determination:} Determines whether a PNE exists in time polynomial in the input size $n$.
        \item \textbf{Efficient Approximation:} If a PNE exists, computes an $\varepsilon$-PNE in time polynomial in $n$ and $\log(1/\varepsilon)$.
    \end{enumerate}
\end{theorem}

We prove Theorem \ref{the2} by constructing a \textit{polynomial-time algorithm} (Algorithm \ref{polyalg}). The key idea behind our approach is to enumerate all possible active sets that include medium-elasticity contestants. For each candidate active set, we perform a binary search over the corresponding feasible range of aggregate production $A$ and verify whether the resulting solution satisfies the PNE conditions. If no candidate active set leads to a valid PNE, we conclude that no PNE exists. Full details are provided in Appendix~\ref{tthe2}.

\section{The Intractable Regime: Hardness and Approximation Limits}
\label{sec:hard}

In this section, we address the general case where the number of medium-elasticity contestants $m$ scales beyond $O(\log n)$. In this regime, the brute force enumeration of active sets becomes prohibitive ($2^m$ grows exponentially), and the interplay between the continuous aggregation $A$ and the discrete participation choices creates a formidable computational barrier. 

This combinatorial explosion raises fundamental questions regarding the computational limits of this regime. Can we design an efficient algorithm that avoids exhaustive search? If an exact solution is out of reach, can we at least approximate it with arbitrary precision at negligible cost?

We answer these questions by establishing a hierarchy of hardness results that strictly bound the attainable efficiency. Formally, we characterize this complexity landscape through three main results:

\begin{enumerate}
    \item \textbf{Exact Hardness (Section \ref{sec:hardness}):} We first prove that the fundamental problem of determining the existence of a PNE in this regime is \emph{NP-complete}, ruling out polynomial-time exact algorithms.
    
    \item \textbf{Inapproximability (Section \ref{sec:hard_approx}):} We strengthen this result by showing that even approximation is hard if the required precision is too high. Specifically, we prove that no algorithm can compute an $\varepsilon$-PNE in time polynomial in $\log(1/\varepsilon)$, unless P=NP.
    
    \item \textbf{Algorithmic Resolution (Section \ref{sec:approx}):} Guided by these negative results, we address the intractability by designing a Fully Polynomial-Time Approximation Scheme (FPTAS). This algorithm achieves the best theoretically possible efficiency—polynomial in $1/\varepsilon$—thereby resolving the computational challenge of the general regime.
\end{enumerate}

\subsection{NP-Completeness of Determining Existence of a PNE for Large $m$}\label{sec:hardness}
We now address the general regime where the number of medium-elasticity contestants, denoted by $m$, is not logarithmically bounded. Specifically, we focus on the case where $m$ is more than $O(\log n)$ (i.e., $m = \Omega(\log n)$). Our main result establishes that the computational tractability collapses in this regime.
\begin{theorem}\label{npc2}
    If there exist more than $\Omega(\log n)$ contestants whose elasticity satisfies $ r_i \in (1, 2] $, determining the existence of a PNE in a Tullock contest is NP-complete.
\end{theorem} 
To prove Theorem \ref{npc2}, we construct a reduction from a variant of the Subset Sum problem.
\begin{definition}[Subset Sum with Large Targets (\texttt{SSLT})]
    Given a set of positive numbers $ \textbf{Z} = \{z_1, \ldots, z_n\} $ and a target $ \bar{z}$ satisfying $\bar{z} \geq 2 \max_{z \in \textbf{Z}} z$, determine whether there exists a subset $ S \subseteq \textbf{Z} $ such that $ \sum_{z \in S} z = \bar{z} $.
\end{definition}

The proof proceeds in three logical steps. First, we establish a structural equivalence between \texttt{SSLT} and Tullock contests (Lemma \ref{npc}). Second, we verify the hardness of the \texttt{SSLT} itself (Lemma \ref{ssbm}). Finally, we synthesize these results to establish that the \emph{general problem} of determining PNE existence with medium elasticity contestants is NP-complete.

\begin{lemma}\label{npc}
    For any instance of \texttt{SSLT}, we can construct in polynomial time a corresponding Tullock contest under medium elasticity $(\forall i, r_i \in (1, 2]) $. \texttt{SSLT} has a solution if and only if the corresponding Tullock contest has a PNE, with the PNE directly corresponding to the solution of the \texttt{SSLT}.
\end{lemma}

Lemma \ref{npc} reveals that the strategic interaction in contests with medium elasticity contestants \emph{effectively encodes} the combinatorial structure of \texttt{SSLT}. Specifically, the binary nature of participation (active vs. inactive) maps to the inclusion decision in the subset sum, while the PNE market-clearing condition imposes the exact constraint required by the target sum. This mapping ensures that the PNE problem inherits the computational hardness of \texttt{SSLT}. Full construction details are provided in Appendix~\ref{D}.

\begin{lemma}\label{ssbm}
    \texttt{SSLT} is an NP-complete problem.
\end{lemma}

We formally establish the hardness of the \texttt{SSLT} variant via a reduction from the standard Subset Sum problem, confirming that the "large target" constraint does not simplify the complexity. The full reduction is provided in Appendix~\ref{C}.

Combining Lemmas \ref{npc} and \ref{ssbm}, we conclude that determining the existence of a PNE is NP-complete for the case where all contestants have medium elasticity. 

Crucially, this result establishes a sharp complexity transition based on the population of medium-elasticity contestants. While the problem is tractable when $m = O(\log n)$ (as shown in Section \ref{sec:unified_algo}), Theorem \ref{npc2} demonstrates that this tractability collapses once the number of medium-elasticity contestants exceeds logarithmic growth ($m \ge \Omega(\log n)$). 

This hardness result is established via a \textit{padding argument} (detailed in Appendix \ref{E}). Specifically, we show that any "hard" instance consisting of $m$ medium-elasticity contestants can be embedded into a larger contest of $n$ players by adding $n-m$ "dummy" contestants with negligible impact (e.g., extremely low efficiency). Consequently, determining the existence of a PNE remains NP-complete for any regime where $m$ is sufficiently large relative to $n$, specifically when $m = \Omega(\log n)$.

\subsection{Hardness of High-Precision Approximation}\label{sec:hard_approx}

Recall that in the efficient regime (Section \ref{sec:unified_algo}), we successfully established an algorithm with runtime polynomial in $(n,\log(1/\varepsilon))$. A natural question is whether this ``high-precision'' efficiency extends to the general regime. While we have proven that determining the \textit{exact} existence of a PNE is NP-complete, this does not theoretically preclude the existence of an efficient approximation scheme. One might hope to design an algorithm that computes an $\varepsilon$-PNE in time polynomial in $(n,\log(1/\varepsilon))$, thereby circumventing the exact hardness while providing solutions with arbitrary precision at negligible cost.

In this subsection, however, we prove that such efficiency is unattainable for the general problem. We demonstrate that the NP-completeness of the existence problem imposes a strict limit on the approximation rate. This limitation stems from a \textit{discrete gap} inherent to the combinatorial reduction: since the underlying problem involves discrete choices, ``No'' instances are not arbitrarily close to ``Yes'' instances but are separated by a minimum non-zero margin determined by the input encoding size. The following lemma formalizes this by establishing a lower bound on the market-clearing error for instances without a PNE.

\begin{lemma}[Gap Lemma]\label{lem:gap}
Let $\mathcal{G}$ be an instance of the Tullock contest constructed via reduction from an \texttt{SSLT} instance with rational inputs. If $\mathcal{G}$ does not admit a PNE, then for any strategy profile where the active set corresponds to a candidate subset and the aggregate production is fixed at the target, the deviation from the market-clearing condition is lower-bounded by:
$$\left| \sum_{i \in \mathcal{I}^A} \sigma_i - 1 \right| \ge \Delta_{\min},$$
where $\Delta_{\min} = 2^{-poly(n)}$ is a value exponentially small in the input size $n$.
\end{lemma}

The significance of Lemma \ref{lem:gap} is that it establishes a "resolution limit." The detailed proof is provided in Appendix \ref{sec:proof_inapprox}. If an algorithm could approximate the PNE with precision $\varepsilon < \Delta_{\min}/2$ in time polynomial in $n$ and $\log(1/\varepsilon)$, it would be capable of distinguishing between instances with zero error (PNE exists) and instances with error at least $\Delta_{\min}$ (no PNE exists). This capability would imply a polynomial-time solution for the NP-complete problem. This reasoning leads directly to our main inapproximability result, whose formal proof is provided in Appendix \ref{sec:proof_inapprox}.

\begin{theorem}[Inapproximability]\label{thm:inapprox}Unless $P = NP$, there is no algorithm that computes an $\varepsilon$-PNE for the general Tullock contest in time polynomial in $n$ and $\log(1/\varepsilon)$.
\end{theorem}

\paragraph{Implication for Algorithm Design.}
This negative result establishes a fundamental barrier: no efficient algorithm can achieve a runtime dependence on precision logarithmic in $1/\varepsilon$ (i.e., polynomial in the number of bits of precision). This implies that the polynomial dependence on $1/\varepsilon$ exhibited by our FPTAS (Section \ref{sec:approx}) cannot be fundamentally improved. Therefore, our algorithm achieves the standard benchmark for efficiency in approximation schemes (FPTAS) while respecting the hardness limits imposed by the NP-complete existence problem.

\subsection{An FPTAS for Finding an $\varepsilon$-PNE}\label{sec:approx}

The complexity results in Section \ref{sec:hardness} establish a tight boundary for the general Tullock contest: finding an exact PNE is NP-complete, and high-precision approximation (logarithmic in $1/\varepsilon$) is equally intractable. Consequently, the most efficient algorithm one can hope for is a Fully Polynomial-Time Approximation Scheme (FPTAS) with dependence polynomial in $1/\varepsilon$. In this section, we present an algorithm that achieves exactly this efficiency bound.

Our goal is to compute an $\varepsilon$-PNE, a standard solution concept where no contestant can significantly gain by deviating. However, directly optimizing for utility deviations is difficult. Instead, we exploit the structural insight from our hardness reduction: the intractability stems specifically from the \textit{exact} market clearing condition ($\sum \sigma_i = 1$). This motivates us to introduce an algorithmic proxy for $\varepsilon$-PNE: the \emph{$\varepsilon$-approximate solution}. This concept relaxes the strict summation constraint to a neighborhood requirement, transforming the intractable exact subset sum problem into a tractable approximate one.

In this subsection, we present our FPTAS, which computes an $\varepsilon$-approximate solution, and prove that such a solution guarantees an $L\varepsilon$-PNE via a Lipschitz continuity constant $L$ related to the Tullock contest instance. While the $\varepsilon$-PNE is defined on payoffs, our algorithm targets the following structural relaxation:

\begin{definition}[$\varepsilon$-approximate solution]\label{def:eps_solution}
A triplet $\{A^*, \mathcal{I}^A, \bm{\sigma^*}\}$ constitutes an $\varepsilon$-approximate solution if and only if the following conditions are satisfied:
\begin{itemize}
    \item Conditions (1), (2) and (3) are identical to those in Proposition~\ref{PNEProp}.
    \item \textbf{Market Clearing:} 
    The individual shares of active contestants falls into an $\varepsilon$ neighborhood of $1$:
    \begin{equation} \label{eq:clear}
        \sum_{i \in \mathcal{I}^A} \sigma_i^* =  (1-\varepsilon,1+\varepsilon) \quad \left( \iff \sum_{i \in \mathcal{I}^A} k(A^*; a_i, r_i) =  (1-\varepsilon,1+\varepsilon) \right).
    \end{equation}
\end{itemize}
\end{definition}

The significance of Definition \ref{def:eps_solution} lies in how it strategically handles the problem's complexity. By enforcing strict best-response consistency (Conditions 1--3), we ensure that any solution remains strict best-response consistency for every contestant. The relaxation is applied exclusively to the \emph{Market Clearing} condition. This is critical because, as shown in Section \ref{sec:hardness}, the computational intractability stems entirely from the combinatorial requirement to hit the aggregate target \textit{exactly}. By relaxing this equality to an interval constraint $\sum \sigma_i \in (1-\varepsilon, 1+\varepsilon)$, we effectively transform the underlying hard problem (Exact Subset Sum) into a tractable variant (Approximate Subset Sum) (Algorithm \ref{subset}), providing the necessary structure for the FPTAS developed in the next subsection.

We now present our main result in this section: an efficient algorithm to compute an $\varepsilon$-PNE in the intractable regime. Crucially, to guarantee a \textit{Fully} Polynomial-Time Approximation Scheme (FPTAS), we require that the elasticity parameters for medium-elasticity contestants (who satisfy $r_i > 1$ by definition) are strictly bounded away from $1$ (i.e., $r_i \ge 1+\delta$ for some constant $\delta > 0$). This assumption is necessary because the Lipschitz constant $L$, which relates our $\varepsilon$-approximate solution to the $\varepsilon$-PNE, scales with $(r_i - 1)^{-1}$. Consequently, to guarantee a final $\varepsilon$-PNE, the algorithm must be executed with a precision of $\varepsilon/L$. This implies that the effective runtime complexity depends on the specific \textit{values} of the input (via the inverse distance to 1) rather than just the \textit{input size} (bit length), rendering the algorithm a \textit{Pseudo-Polynomial Time Approximation Scheme} (PPTAS). While a PPTAS is also a widely acceptable efficiency benchmark (detailed in Appendix \ref{sec:pptas_discussion}), the strict bound $\delta$ allows us to bound $L$ by a constant, thereby achieving the stronger FPTAS guarantee stated in the following Theorem.

\begin{theorem}[Efficiency and Correctness of FPTAS]\label{thm:fptas_main}
    If a PNE exists, there exists an algorithm to compute an $\varepsilon$-approximate solution in polynomial time with respect to $n$ and $1/\varepsilon$. Furthermore, this output constitutes an $(L\varepsilon)$-PNE, where $L$ is a bounded constant depending on the contest parameters $\bm{a}$ and $\bm{r}$.
\end{theorem}

We prove Theorem \ref{thm:fptas_main} by explicitly constructing such an algorithm (\textbf{Algorithm \ref{alg:FPTAS}}) and analyze its properties. The algorithm circumvents the combinatorial barrier by combining a geometric discretization of the aggregate production $A$ with a polynomial-time approximation scheme for the Subset Sum problem.

\begin{algorithm}[H]
\SetAlgoNoEnd
\caption{FPTAS for $\varepsilon$-PNE in Tullock Contests}
\label{alg:FPTAS}
\KwIn{Number of contestants $n$, efficiency vector $\bm{a}$, elasticity vector $\bm{r}$, approximation parameter $\varepsilon$.}
\KwOut{A set of $\varepsilon$-PNE $Y$ (if found), otherwise ``No solution''.}
\BlankLine

\textbf{Step 1: Partition the Space of Aggregate production $A$} \\
\quad Compute $\underline{A} = \min_{i \in \mathcal{I}^2} \underline{A}_i$, $\bar{A} = \max_{i \in \mathcal{I}^2}\bar{A}_i$ and $A_{\max}$ according to Eq. \eqref{eq:thresholds}\;
\quad Divide the space of $A$ into three intervals: $(0, \underline{A}], (\underline{A}, \bar{A}], (\bar{A}, A_{\max})$\\
\quad Initialize an empty solution set $Y \gets \emptyset$\;
\textbf{Step 2: Solve for $(0, \underline{A}]$ and $(\bar{A}, A_{\max})$ Using Binary Search} \\

\quad Perform binary search for $A$ in two separate ranges:
\begin{itemize}
    \item \textit{For $(0, \underline{A}]$}: directly apply binary search to find $A^*$ such that $\sum \sigma_i^* = 1$.
    \item \textit{For $(\bar{A}, A_{\max})$}: apply binary search to find $A^*$ on contestants with $r \leq 1$ such that $\sum \sigma_i^* = 1$.
\end{itemize}
\quad Add $(A^*, \bm{\sigma^*}, \mathcal{I}^A)$ to $Y$ whenever a valid PNE is found in either range.

\textbf{Step 3: Solve for $(\underline{A}, \bar{A}]$ via discretization and \texttt{APPROX-SUBSET-SUM}.}\\

\quad Construct a discretized grid
\[
\mathcal{A}_1 \gets
\left\{ A \,\middle|\,
A = \underline{A}\cdot
\left(1 + \min\{\underline{r}^2,\underline{r}'(\underline{r}'-1)\}\cdot\frac{\varepsilon}{n}\right)^k,
A \leq \overline{A},
\;
k \in \mathbb{N}^+
\right\}
\]

\quad Compute $ \{\underline{A_i},\bar{A_i}\}_{i\in \mathcal{I}^2} $ and let
$ \mathcal{A}_2 \gets \bigcup_{i\in \mathcal{I}^2} \{\underline{A_i},\bar{A_i}\} $\;

\quad and define the final node set
$ \mathcal{A} \gets \mathcal{A}_1 \cup \mathcal{A}_2 $\;

\quad For each candidate $A \in \mathcal{A}$\\
\begin{itemize}
    \item Compute best-response action shares $\sigma_i(A)$ for all contestants $i\in\mathcal{I}$\;
    \item $S_0 \gets \sum_{i:\, \bar{A}_i<A<\underline{A}_i \text{ and } i\in\mathcal{I}^1} \sigma_i(A)$ \quad and \quad $\bm{\sigma}(A) \gets \bigl(\sigma_i(A)\bigr)_{i:\, A\in[\underline{A}_i,\bar{A}_i]}$
    \item Run \texttt{APPROX-SUBSET-SUM}$(S_0,\bm{\sigma}(A),\varepsilon)$ to verify if $A$ constitutes an approximate solution, and add $(A^*, \bm{\sigma^*}, \mathcal{I}^A)$ to $Y$ if \texttt{TRUE}.
\end{itemize}
\textbf{Step 4: Verify and Return the Solution Set} \\
   \quad For each $(A^*, \bm{\sigma^*}, \mathcal{I}^A) \in Y$, add $(A^*, \bm{\sigma^*}, \mathcal{I}^A)$ to $Y_{\text{valid}}$ if it satisfies $\varepsilon$-PNE conditions.\\
\Return $Y_{\text{valid}}$ (set of verified $\varepsilon$-PNEs) if $Y_{\text{valid}} \neq \emptyset$ else ``No valid $\varepsilon$-PNE found''.\;
\end{algorithm}

Here are some remarks about the operational Logic of Algorithm \ref{alg:FPTAS}. In Step 1, the algorithm partitions the search space based on the stability of the active set. Then in Step 2, we use binary search to efficiently identify the PNEs in the boundary intervals $(0, \underline{A}]$ and $(\bar{A}, A_{\max})$, as the corresponding set of active contestants under a PNE is uniquely determined when either all contestants or only contestants with $r\le 1$ are active, due to the strict monotonicity of best responses in these regions.

The core complexity lies in Step 3, in which we need to pin down $A^*$ in the medium elasticity regime $(\underline{A}, \bar{A}]$. Here, the active set depends combinatorially on the selection of contestants. To resolve the continuous nature of $A$ in this region, we construct a discrete candidate set $\mathcal{A} = \text{sorted}(\bm{A}_1 \cup \bm{A}_2)$. This set integrates:
\begin{itemize}[leftmargin=10pt]
    \item $\bm{A}_1$: Nodes evenly spaced across $(\underline{A}, \bar{A}]$, formally defined as\\
    $
    \bm{A}_1 = \{A \mid A = \underline{A}\left(1 + \min\{\underline{r}^2,\underline{r}'(\underline{r}'-1)\} \cdot \frac{\varepsilon}{n}\right)^k, A \leq \bar{A}, k \in \mathbb{N}^+\}
    $.
    where $\underline{r}$ denotes the minimum value of $r$ in the contest instance, and $\underline{r}'$ represents the minimum value of $r$ in $\mathcal{I}^2$.
    \item $\bm{A}_2$: The discontinuity points $\{\underline{A}_i, \bar{A}_i\}$ where contestant behavior changes abruptly.
    
\end{itemize}

The sufficiency of this discretization is guaranteed by the following lemma, which ensures that searching over $\mathcal{A}$ is equivalent to searching the continuous interval up to $\varepsilon$-precision.

\begin{lemma}\label{select}
    For any $\varepsilon > 0$, if there exists an $\varepsilon$-approximate solution in an interval $[A_1, A_2]$, where $A_1$ and $A_2$ are two adjacent nodes in $\mathcal{A}$, then at least one of these nodes, $A_1$ or $A_2$, constitutes an $\varepsilon$-approximate solution.
\end{lemma}

The proof is straightforward and deferred in Appendix \ref{f}. Lemma \ref{select} validates our discretization strategy in Step 3 of Algorithm \ref{alg:FPTAS} by ensuring that any valid solution in the continuous domain has a representative within the generated grid $\mathcal{A}$. Therefore, we can reduce the continuous search space to a finite, polynomial-sized set of candidates to ensure the running time efficiency without sacrificing approximation accuracy.

\paragraph{Proof sketch of Theorem \ref{thm:fptas_main}.}
With the continuous search space effectively reduced to the discrete set $\mathcal{A}$, we can formally establish the time complexity. The runtime analysis distinguishes between the boundary intervals and the ambiguous interval:

\begin{itemize}
    \item \textbf{Boundary Intervals (Binary Search):} 
    For $(0, \underline{A}]$ and $(\bar{A}, A_{\max})$, the active set is fixed. As detailed in Appendix F, a binary search to precision $\Theta(\varepsilon/nL)$ requires $O(\log \frac{\bar{A}n}{\varepsilon})$ iterations. This logarithmic cost is negligible compared to the main loop.

    \item \textbf{Ambiguous Interval (Discretization + Subset Sum):}
    The computational bottleneck lies in the interval $(\underline{A}, \bar{A}]$. The number of grid points is bounded by $|\mathcal{A}| = O(\frac{n}{\varepsilon})$. For each node $A \in \mathcal{A}$:
    \begin{enumerate}
        \item Computing best-response shares for $n$ contestants takes $O(n \log \frac{n}{\varepsilon})$.
        \item Executing the \texttt{APPROX-SUBSET-SUM} subroutine (Algorithm \ref{subset}) takes $O(\frac{n^3}{\varepsilon} \log^2 n)$.
    \end{enumerate}
    The total complexity is dominated by the product of the grid size and the subset sum routine:
    \[
    O\left(\frac{n}{\varepsilon}\right) \times O\left(\frac{n^3}{\varepsilon} \log^2 n\right) = O\left(\frac{n^4}{\varepsilon^2} \log^2 n\right).
    \]
\end{itemize}
This confirms that Algorithm \ref{alg:FPTAS} runs in polynomial time with respect to $n$ and $1/\varepsilon$. Full detailed proof is shown in Appendix~\ref{gg}.

Having established the algorithmic efficiency, we now address the second claim of Theorem \ref{thm:fptas_main}: that the output constitutes an $(L\varepsilon)$-PNE. This requires establishing the equivalence between $\varepsilon$-approximate solution defined by the \textit{approximate market clearing} condition and $\varepsilon$-PNE defined by \textit{individual optimality}. We show that this equivalence follows from the Lipschitz continuity of several functions derived from the utility function, which can be established through a careful analysis of the best-response mapping. The complete argument is deferred to Appendix~\ref{g}.

\section{Experiment and Analysis}

In this section, we present computational results evaluating the performance and structural properties of the computed $\varepsilon$-PNE. Our focus is twofold: we first analyze the runtime trends for varying contest parameters, and then investigate the structure of equilibrium solutions.
\vspace{-2mm}
\subsection{Game Instance Generation and Implementation of Algorithm \ref{alg:FPTAS}}
To evaluate the scalability and robustness of our FPTAS, we generate synthetic Tullock contest instances by varying the number of contestants $n$ and sampling the contest parameters from specific distributions. We categorize the test instances into four distinct types, distinguished by the distribution of the efficiency parameter $a$ and the magnitude of the reward $R$. Specifically, the types are defined as follows: Type 1 consists of instances where $a$ is drawn from parameters $(1, 0.25)$ with a reward $R=10$; Type 2 draws $a$ from the interval $(2.5, 5)$ with $R=10$; Type 3 draws $a$ from $(1.5, 2.5)$ with a larger reward $R=20$; and Type 4 draws $a$ from $(2.5, 5)$ with $R=20$. For all instances, the elasticity parameters $r_i$ are generated to ensure a significant presence of medium-elasticity contestants ($r_i \in (1, 2]$), thereby placing the problem in the intractable regime and triggering the core logic of our approximation scheme.

Our implementation include several core Python modules including binary search, MAMD algorithm and a solver for the subset sum problems. The code is anonymously available in the supplementary documents. One notable challenge in the implementation was managing the trade-off between accuracy and computational efficiency. The use of the trimmed subset sum algorithm ensures scalability, but further improvements in this step remain an open question, as discussed in Section \ref{dis}.
\vspace{-2mm}
\subsection{Runtime Analysis}
\begin{figure}[H]
    \centering
    \begin{subfigure}[b]{0.49\linewidth}
        \includegraphics[width=\linewidth]{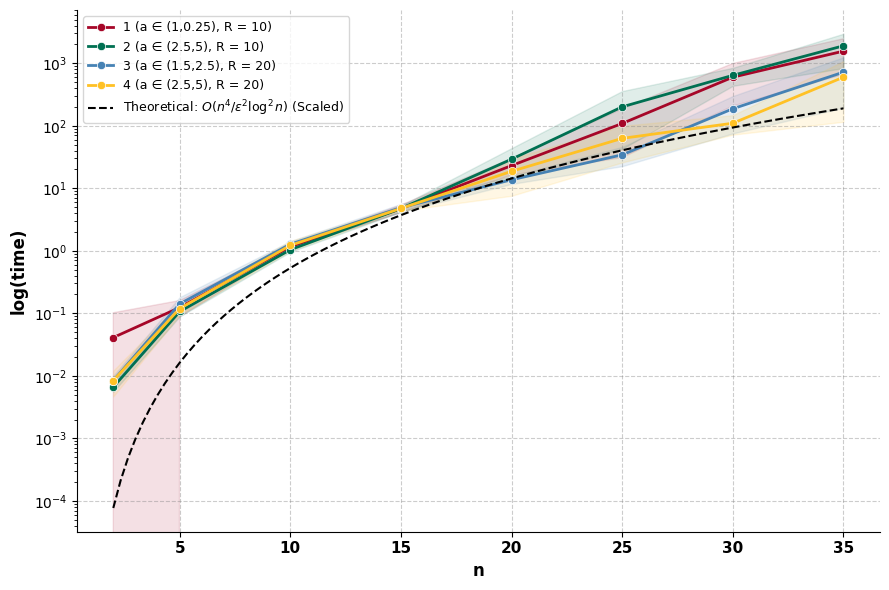}
        \caption{Runtime of our FPTAS across different instances with various sizes. Error bars represents thestandard deviation over 5 randomly sampled instances.}
        \label{fig:runtime_analysis}
    \end{subfigure}
    \hfill
    \begin{subfigure}[b]{0.49\linewidth}
        \includegraphics[width=\linewidth]{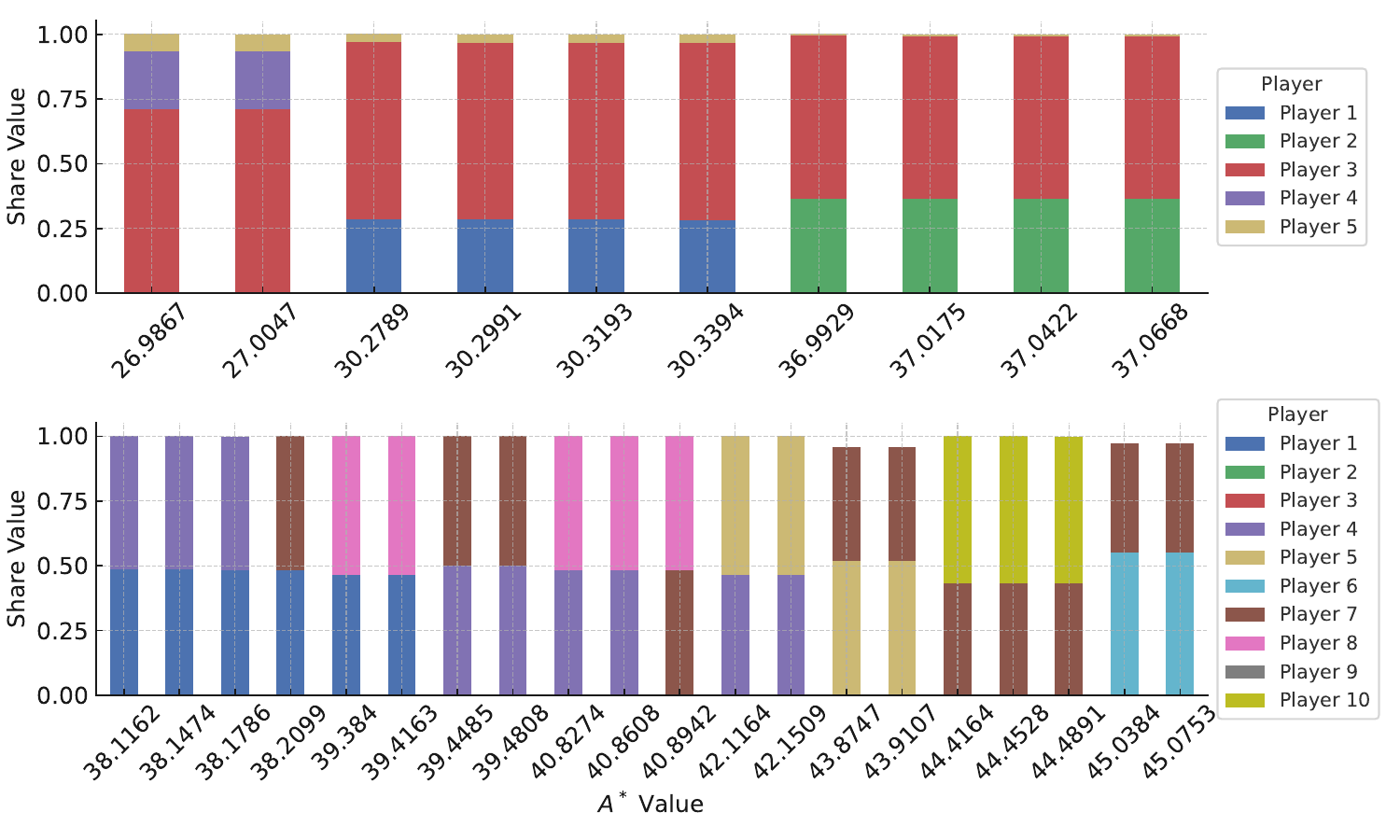}
    \caption{The structures of all $\varepsilon$-PNE solutions found by our FPTAS. Each bar represents the action shares of all active players under different values of $A^*$. }
    \label{fig:effort_distribution}
    \end{subfigure}
\end{figure}

In this subsection, we analyze the computational performance of Algorithm \ref{alg:FPTAS} by measuring the runtime required to compute $\varepsilon$-PNE for our constructed Tullock contest instances. The results, averaged over five runs for each instance type, are summarized in Figure~\ref{fig:runtime_analysis}.

As illustrated in Figure \ref{fig:runtime_analysis}, the empirical runtime trajectory aligns remarkably well with the $O(\frac{n^4}{\varepsilon^2}\log^2 n)$ complexity established in Theorem \ref{thm:fptas_main}. The polynomial growth observed in our experiments precisely mirrors the theoretical results, providing strong empirical validation for our complexity analysis. The error bars shows that runtime can vary a lot across different instances, especially when $n$ gets larger. This suggests that the parameter configurations in Tullock contests play a key role in determining their computational efficiency, as our key finding in Theorem \ref{thm:fptas_main} suggests, the runtime not only depend on $n,\varepsilon$ but a problem-specific constant $L$.
\vspace{-5mm}

\vspace{-2mm}
\subsection{Analysis of $\varepsilon$-PNE Structure}

Our equilibrium finding approach facilitates us to further investigate the nature of the computed $\varepsilon$-PNE. We analyzed the composition of different equilibrium solutions and illustrate some key observations using an example Tullock contest instance with $n=10$ contestants. Figure~\ref{fig:effort_distribution} presents a stacked area plot illustrating the effort contributions of active players for different values of $A^*$. The choice of a stacked area plot allows us to observe both individual player contributions and overall trends across different equilibrium states.
\begin{itemize}[leftmargin=10pt]
    \item \textbf{Diversity of Equilibria:} The solutions exhibit substantial variation in active player composition even for closely related values of $A^*$. For example, in the second diagram of Fig \ref{fig:effort_distribution}, for $A^* \approx 39.8$, the active player set transitions from \{1, 8\} to \{4,7\}, illustrating the flexibility of $\varepsilon$-PNE solutions.
    \item \textbf{Localized Stability:} In some regions of $A^*$, the set of active players remains stable over a range of values. For example, in the first diagram of Fig \ref{fig:effort_distribution}, between $A^* \approx 26$ and $A^* \approx 27$, Players 3,4 and 5 consistently maintain dominance, suggesting equilibrium stability within this range.
\end{itemize}

These findings highlight the inherent flexibility and complexity of $\varepsilon$-PNE in Tullock contests. Unlike exact PNE, where equilibrium uniqueness may be possible under certain conditions, $\varepsilon$-PNE solutions demonstrate a range of viable strategic adaptations. The presence of discrete shifts suggests that equilibrium selection is highly sensitive to small perturbations in effort distributions, potentially reflecting real-world decision-making scenarios where agents react adaptively to competition dynamics. Overall, our observations emphasize the need for a robust computational framework for exploring $\varepsilon$-PNE landscapes, since diverse equilibrium configurations can exist even within a narrow parameter space. Moreover, our FPTAS, given by Algorithm \ref{alg:FPTAS}, provides an effective tool to systematically uncover and analyze the complex structures of these $\varepsilon$-PNEs. 
\vspace{-2mm}
\section{Discussion}\label{dis}

This paper explores the computational complexity of computing PNEs in general Tullock contests by identifying the structural determinants of tractability. Our primary contributions include:

\begin{itemize}[leftmargin=10pt]
    \item \textbf{Complexity Phase Transition:} We established that the computational complexity of determining the existence of a PNE and computing it is governed by the number of contestants $m$ whose elasticity parameters fall within the medium regime $(1, 2]$. We identified a sharp transition: when $m = O(\log n)$ (\textit{Easy Case}), the problem is efficiently solvable. However, when $m$ is beyond the logarithmic scale (i.e, $m = \omega(\log n)$) (\textit{Hard Case}), determining the existence of a PNE becomes NP-complete.

    \item \textbf{Convergence Rate Analysis:} We provided a refined understanding of the dependence on precision $\varepsilon$. In the Easy Case, we developed algorithms that compute an $\varepsilon$-PNE in time \emph{polynomial in $n$ and $\log(1/\varepsilon)$}, achieving high-precision solutions rapidly. Conversely, for the Hard Case, we proved that such logarithmic dependence is unattainable unless P=NP, and we instead developed an FPTAS with a runtime polynomial in $n$ and $1/\varepsilon$.
    
    \item \textbf{Empirical Validation:} We implemented our proposed algorithms in a Python module and evaluated their performance across diverse contest configurations. Our experimental results exhibit a striking alignment with our theoretical complexity analysis, particularly the $O(\frac{n^4}{\varepsilon^2}\log^2 n)$ bound for the FPTAS, confirming the predicted shift in computational effort between the tractable and intractable regimes.
\end{itemize}

Our results have both theoretical and practical significance. Theoretically, our work advances the understanding of how elasticity conditions and the scale of heterogeneity influence the tractability of PNEs. The distinction between $\log(1/\varepsilon)$ and $1/\varepsilon$ dependence highlights the intrinsic difficulty of the Hard Case and underscores the necessity of approximation techniques. Practically, our algorithms provide robust tools for analyzing strategic interactions in complex systems, such as decentralized resource allocation in blockchains.

An important open question remains: can the FPTAS be further optimized to achieve improved efficiency? A significant portion of the current complexity arises from the \textit{Merge-and-Trim Approximation} used to handle real-number subset sum problems. While our current implementation matches the $1/\varepsilon^2$ theoretical growth perfectly, investigating alternative, faster approximation techniques for the summation steps remains an open challenge. Improving this component could further enhance the scalability of computing $\varepsilon$-PNE solutions in large-scale asymmetric contests.

\bibliographystyle{plainnat}
\bibliography{citation} 

@String{Computing = "Computing" }

@String{Computer = "{IEEE} Computer" }

@String{Springer = "Springer-Verlag" }

@article{Tullock1980,
  title={Efficient Rent-Seeking},
  author={Tullock, Gordon},
  year={1980},
  publisher={Toward a Theory of the Rent-Seeking Society}
}

@article{skaperdas1996contest,
  title={Contest success functions},
  author={Skaperdas, Stergios},
  journal={Economic theory},
  volume={7},
  pages={283--290},
  year={1996},
  publisher={Springer}
}

@book{luce1959individual,
  title={Individual choice behavior},
  author={Luce, R Duncan},
  volume={4},
  year={1959},
  publisher={Wiley New York}
}

@article{baye2003strategic,
  title={The strategic equivalence of rent-seeking, innovation, and patent-race games},
  author={Baye, Michael R and Hoppe, Heidrun C},
  journal={Games and economic behavior},
  volume={44},
  number={2},
  pages={217--226},
  year={2003},
  publisher={Elsevier}
}

@article{baye1993rigging,
  title={Rigging the lobbying process: An application of the all-pay auction},
  author={Baye, Michael R and Kovenock, Dan and De Vries, Casper G},
  journal={The American Economic Review},
  volume={83},
  number={1},
  pages={289},
  year={1993},
  publisher={American Economic Association}
}

@inproceedings{deng2024competition,
  title={Competition among pairwise lottery contests},
  author={Deng, Xiaotie and Gan, Hangxin and Li, Ningyuan and Li, Weian and Qi, Qi},
  booktitle={Proceedings of the AAAI Conference on Artificial Intelligence},
  volume={38},
  number={9},
  pages={9662--9669},
  year={2024}
}

@article{clark1998contest,
  title={Contest success functions: an extension},
  author={Clark, Derek J and Riis, Christian},
  journal={Economic Theory},
  volume={11},
  pages={201--204},
  year={1998},
  publisher={Springer}
}

@article{Cong2018,
  title={Decentralized mining in centralized pools},
  author={Cong, Lin William and He, Zhiguo and Li, Jiasun},
  journal={The Review of Financial Studies},
  year={2018}
}

@inproceedings{chen2019axiomatic,
  title={An axiomatic approach to block rewards},
  author={Chen, Xi and Papadimitriou, Christos and Roughgarden, Tim},
  booktitle={Proceedings of the 1st ACM Conference on Advances in Financial Technologies},
  pages={124--131},
  year={2019}
}

@article{leshno2020bitcoin,
  title={Bitcoin: An axiomatic approach and an impossibility theorem},
  author={Leshno, Jacob D and Strack, Philipp},
  journal={American Economic Review: Insights},
  volume={2},
  number={3},
  pages={269--286},
  year={2020},
  publisher={American Economic Association 2014 Broadway, Suite 305, Nashville, TN 37203}
}

@inproceedings{cavallo2013winner,
  title={Winner-take-all crowdsourcing contests with stochastic production},
  author={Cavallo, Ruggiero and Jain, Shaili},
  booktitle={Proceedings of the AAAI Conference on Human Computation and Crowdsourcing},
  volume={1},
  pages={34--41},
  year={2013}
}

@inproceedings{rokicki2014competitive,
  title={Competitive game designs for improving the cost effectiveness of crowdsourcing},
  author={Rokicki, Markus and Chelaru, Sergiu and Zerr, Sergej and Siersdorfer, Stefan},
  booktitle={Proceedings of the 23rd ACM International Conference on Conference on Information and Knowledge Management},
  pages={1469--1478},
  year={2014}
}

@article{arnosti2022bitcoin,
  title={Bitcoin: A natural oligopoly},
  author={Arnosti, Nick and Weinberg, S Matthew},
  journal={Management Science},
  volume={68},
  number={7},
  pages={4755--4771},
  year={2022},
  publisher={INFORMS}
}

@article{Thum2018,
  title={The economic cost of Bitcoin mining},
  author={Thum, Marcel},
  journal={CESifo Forum},
  year={2018}
}

@article{Easley2019,
  title={From mining to markets: The evolution of Bitcoin transaction fees},
  author={Easley, David and O'Hara, Maureen and Basu, Soumya},
  journal={Journal of Financial Economics},
  year={2019}
}

@article{Cornes2005,
  title={Asymmetric contests with general technologies},
  author={Cornes, Richard and Hartley, Roger},
  journal={Economic Theory},
  volume={26},
  number={4},
  pages={923--946},
  year={2005},
  publisher={Springer}
}

@inproceedings{yao2024human,
  title={Human vs. Generative AI in Content Creation Competition: Symbiosis or Conflict?},
  author={Yao, Fan and Li, Chuanhao and Nekipelov, Denis and Wang, Hongning and Xu, Haifeng},
  booktitle={International Conference on Machine Learning},
  pages={235:56885-56913},
  year={2024},
  organization={PMLR}
}

@book{Mart1990,
author = {Martello, Silvano and Toth, Paolo},
title = {Knapsack problems: algorithms and computer implementations},
year = {1990},
isbn = {0471924202},
publisher = {John Wiley \& Sons, Inc.},
address = {USA}
}

@article{rosen1965existence,
  title={Existence and uniqueness of equilibrium points for concave n-person games},
  author={Rosen, J Ben},
  journal={Econometrica: Journal of the Econometric Society},
  pages={520--534},
  year={1965},
  publisher={JSTOR}
}

@article{bravo2018bandit,
  title={Bandit learning in concave N-person games},
  author={Bravo, Mario and Leslie, David and Mertikopoulos, Panayotis},
  journal={Advances in Neural Information Processing Systems},
  volume={31},
  year={2018}
}

@inproceedings{even2009convergence,
  title={On the convergence of regret minimization dynamics in concave games},
  author={Even-Dar, Eyal and Mansour, Yishay and Nadav, Uri},
  booktitle={Proceedings of the forty-first annual ACM symposium on Theory of computing},
  pages={523--532},
  year={2009}
}

@inproceedings{Elkind2024,
  title={Continuous-Time Best-Response and Related Dynamics in Tullock Contests with Convex Costs},
  author={Elkind, Edith and Ghosh, Abheek and Goldberg, Paul W.},
  booktitle={Proceedings of the 20th Conference on Web and Internet Economics (WINE)},
  year={2024},
  note={arXiv preprint arXiv:2402.08541}
}

@article{ghosh2023best,
  title={Best-response dynamics in tullock contests with convex costs},
  author={Ghosh, Abheek},
  journal={arXiv preprint arXiv:2310.03528},
  year={2023}
}

@inproceedings{Ghosh2023,
  title={Best-Response Dynamics in Lottery Contests},
  author={Ghosh, Abheek and Goldberg, Paul W},
  booktitle={Proceedings of the 24th ACM Conference on Economics and Computation},
  pages={736--736},
  year={2023}
}

@article{bryan2022r,
  title={R\&D competition and the direction of innovation},
  author={Bryan, Kevin A and Lemus, Jorge and Marshall, Guillermo},
  journal={International Journal of Industrial Organization},
  volume={82},
  pages={102841},
  year={2022},
  publisher={Elsevier}
}

@misc{DemandSage2024,
  title={Blockchain Statistics 2024},
  author={DemandSage},
  year={2024},
  note={\url{https://www.demandsage.com/blockchain-statistics/}}
}

@misc{IEEEBlockchain2019,
  title={Blockchain Technology: Prospects, Challenges, and Opportunities},
  author={IEEE Blockchain Initiative},
  year={2019},
  note={\url{https://blockchain.ieee.org/technicalbriefs/june-2019/blockchain-technology-prospects-challenges-and-opportunities}}
}

@article{van2013theory,
  title={The theory of contests: A unified model and review of the literature},
  author={Van Long, Ngo},
  journal={European Journal of Political Economy},
  volume={32},
  pages={161--181},
  year={2013},
  publisher={Elsevier}
}

@article{szidarovszky1997existence,
  title={On the existence and uniqueness of pure Nash equilibrium in rent-seeking games},
  author={Szidarovszky, Ferenc and Okuguchi, Koji},
  journal={Games and Economic Behavior},
  volume={18},
  number={1},
  pages={135--140},
  year={1997},
  publisher={Elsevier}
}

@article{reny1999existence,
  title={On the existence of pure and mixed strategy Nash equilibria in discontinuous games},
  author={Reny, Philip J},
  journal={Econometrica},
  volume={67},
  number={5},
  pages={1029--1056},
  year={1999},
  publisher={Wiley Online Library}
}

@article{olszewski2023equilibrium,
  title={Equilibrium existence in games with ties},
  author={Olszewski, Wojciech and Siegel, Ron},
  journal={Theoretical Economics},
  volume={18},
  number={2},
  pages={481--502},
  year={2023},
  publisher={Wiley Online Library}
}

@book{KellererPferschyPisinger2004,
  author    = {Hans Kellerer and Ulrich Pferschy and David Pisinger},
  title     = {Knapsack Problems},
  publisher = {Springer},
  year      = {2004},
  series    = {Springer Series in Operations Research and Financial Engineering},
  isbn      = {978-3-540-24777-7},
  doi       = {10.1007/978-3-540-24777-7}
}

@techreport{ewerhart2024tullock,
  author       = {Ewerhart, Christian},
  title        = {Solving the n-player Tullock contest},
  institution  = {Department of Economics, University of Zurich},
  type         = {Working Paper},
  number       = {447},
  address      = {Zurich, Switzerland},
  year         = {2024},
  doi          = {10.5167/uzh-261058},
  url          = {https://hdl.handle.net/10419/301657}
}

@inproceedings{daskalakis2006note,
  title={A note on approximate Nash equilibria},
  author={Daskalakis, Constantinos and Mehta, Aranyak and Papadimitriou, Christos},
  booktitle={International Workshop on Internet and Network Economics},
  pages={297--306},
  year={2006},
  organization={Springer}
}

@article{fu2016disclosure,
  title={Disclosure policy in Tullock contests with asymmetric stochastic entry},
  author={Fu, Qiang and Lu, Jingfeng and Zhang, Jun},
  journal={Canadian Journal of Economics/Revue canadienne d'{\'e}conomique},
  volume={49},
  number={1},
  pages={52--75},
  year={2016},
  publisher={Wiley Online Library}
}

\newpage
\appendix
\newpage
\begin{center}
    \Large{\textbf{Appendix}} 
\end{center}

\noindent \textbf{Organization of the Appendix}
This appendix provides the detailed proofs, supplementary structural analysis, and algorithmic descriptions omitted from the main text. It is organized as follows:
\begin{itemize}[leftmargin=15pt]
    \item \textbf{Appendix \ref{sec:struct_prop}: Structural Properties.} We establish fundamental properties of the Tullock contest model, including scaling invariance (Lemma \ref{lem:R=1}), equilibrium characterization (Proposition \ref{PNEProp}), and monotonicity of best responses (Proposition \ref{sa}). We also analyze the impossibility of equilibrium in the purely convex regime.
    \item \textbf{Appendix \ref{sec:eff_regime}: Efficient Regimes.} We provide proofs for the polynomial-time solvability of contests in the small elasticity regime and the logarithmic medium elasticity regime (Theorem \ref{the2}). This section also details the approximation guarantees for the Bisection and Mirror Descent algorithms.
    \item \textbf{Appendix \ref{sec:hard_proofs}: Computational Hardness.} We provide the complete reduction showing the NP-completeness of the PNE existence problem, including the hardness of the variant Subset Sum problem (Lemma \ref{ssbm}) and the padding argument for the general theorem (Theorem \ref{npc2}).
    \item \textbf{Appendix \ref{sec:fptas_proofs}: FPTAS Analysis.} This section contains the theoretical foundation for our approximation scheme, including the grid sufficiency lemma (Lemma \ref{select}), the derivation of Lipschitz constants (Lemma \ref{Lip}), and the final proof of correctness for the FPTAS (Theorem \ref{thm:fptas_main}).
    \item \textbf{Appendix \ref{ALG}: Algorithms.} We present the detailed pseudocode for the algorithms discussed in the paper.
\end{itemize}

\section{Structural Properties of Tullock Contests}\label{sec:struct_prop}

In this section, we derive key structural properties of the contest model. These results simplify the PNE analysis and justify the reduction of the strategy space in our algorithms.

\subsection{Scaling Invariance}\label{append:lem:R=1}
First, we show that the PNE structure is invariant to the scaling of the reward $R$ (Proof of Lemma \ref{lem:R=1}). This allows us to normalize $R=1$ in our algorithmic design without loss of generality.

\begin{proof}
Given  $ \mathcal{G} = \{ (a_i, r_i) \}_{i=1}^n \cup \{ R \} $, let $a'_i = a_i R^{r_i}$ for each $i$ and consider the new instance $\mathcal{G}' = \{ (a'_i, r_i) \}_{i=1}^n \cup \{ 1 \}  $. Then we have 
\begin{eqnarray*}
  & &  \boldsymbol{x}^* = (x_1^*,x_2^*,\cdots,x_n^*)\text{  is a PNE in } \mathcal{G}    \\ 
  & \Leftrightarrow  &   \frac{ a_i (x^*_i)^{r_i}}{\sum_{j=1}^n a_j (x^*_j)^{r_j}} \cdot R - x^*_i \geq  \frac{ a_i (x_i)^{r_i}}{ a_i (x_i)^{r_j}  + \sum_{j \not = i}  a_j (x^*_j)^{r_j} } \cdot R - x_i \qquad \forall x_i >0, i \in [n] \\ 
  & \Leftrightarrow  &   \frac{ a_i R^{r_i} (\frac{x^*_i}{R})^{r_i}}{\sum_{j=1}^n a_j R^{r_j} (\frac{x^*_j}{R})^{r_j}} \cdot 1   - \frac{x^*_i}{R} \geq  \frac{ a_i R^{r_i} (\frac{x_i}{R})^{r_i}}{ a_i R^{r_i} (\frac{x_i}{R})^{r_j}  + \sum_{j \not = i}  a_j R^{r_j} (\frac{x^*_j}{R})^{r_j} }  \cdot 1  - \frac{x_i}{R} \qquad \forall x_i >0,  i \in [n] \\ 
   & \Leftrightarrow  &   \frac{ a'_i   (\frac{x^*_i}{R})^{r_i}}{\sum_{j=1}^n a'_j  (\frac{x^*_j}{R})^{r_j}}   - \frac{x^*_i}{R} \geq  \frac{ a'_i  (\frac{x_i}{R})^{r_i}}{ a'_i   (\frac{x_i}{R})^{r_j}  + \sum_{j \not = i}  a'_j   (\frac{x^*_j}{R})^{r_j} }   - \frac{x_i}{R} \qquad \forall x_i >0, i \in [n] \\   
   & \Leftrightarrow  & \boldsymbol{x}^*/R = (x_1^*/R,x_2^*/R,\cdots,x_n^*/R)\text{  is a PNE in } \mathcal{G}'    
\end{eqnarray*} 
\end{proof}
 
\subsection{Equilibrium Characterization}\label{Prof.PNE}
The following proposition provides the necessary and sufficient conditions for a tuple $\{A^*, \mathcal{I}^A, \bm{\sigma^*}\}$ to constitute a PNE (Proof of Proposition \ref{PNEProp}). This characterization forms the basis for the logic of our exact and approximate algorithms.

\begin{proof}
We first show $\Rightarrow$. If $\{A^*, \mathcal{I}^A, \bm{\sigma^*}\}$ constitutes a PNE, then all contestants in $\mathcal{I}^A$ are active, and all contestants not in $\mathcal{I}^A$ are inactive. 

For all contestants in $\mathcal{I}^A$ with $r_i < 1$ are active, it has to be $A^* <  A_{\max}$. For all contestants in $\mathcal{I}^A$ with $r_i = 1$ are active, it has to be $A^*<a_i$. For all contestants in $\mathcal{I}^A$ with $r_i > 1$ are active, it has to be $A^* \leq \bar{A}_i$. Otherwise, there exists at least one active contestant who should not participate. So that for contestants with $ r_i < 1 $, $A^* < A_{\max}$, for contestants with $ r_i = 1 $, $A^*<\min_{i \in \mathcal{I}^A} {a_i}$, and for contestants with $ r_i > 1 $, $A^* \leq \min_{i \in \mathcal{I}^A} \bar{A}_i$. 

For all contestants not in $\mathcal{I}^A$ are inactive with $r_i < 1$, there is no such case since needs $A > A_{max}$, which do not exists since each contestant's effort can not larger than $1$. For all contestants not in $\mathcal{I}^A$ are inactive with $r_i = 1$, it has to be $A^* \geq a_i$. And for all contestants not in $\mathcal{I}^A$ are inactive with $r_i > 1$, it has to be $A^* \geq \underline{A}_i$. Otherwise there exists at least one inactive contestant who wants to participate. So that for contestants with $ r_i = 1 $, $\max_{i \notin \mathcal{I}^A}{a_i}\leq A^*$, and for contestants with $ r_i > 1 $, $\max_{i \notin \mathcal{I}^A} \underline{A_i} \leq A^*$.

If $\mathcal{I}^A$ is the set of active contestants in equilibirum, then it has to be that $\forall i \in \mathcal{I}^A$, $(\sigma_i^*,A^*)$ is consistent with their best responses, i.e, $b(A^*,\sigma_i^*;a_i,r_i) = 0,\forall i\in \mathcal{I}^A$.

Also, for the equilibirum to be feasible, the action shares of all active contestants sum up to 1, that is $\sum_{i\in\mathcal{I}^A}\sigma_i^* = 1$.

Next we show $\Leftarrow$. To show that $\{A^*, \mathcal{I}^A, \bm{\sigma^*}\}$ constitutes a Nash equilibrium, we only need to show that no contestant has incentive to deviate. First, $\forall i \notin \mathcal{I}^A$, given $A^*$, $\sigma_i = 0$ is (one of) the best response, so that inactive contestants have no incentive to participate. Next, $\forall i \in \mathcal{I}^A$,  given $A^*$, since $(A^*,\sigma_i^*;a_i,r_i) = 0$, $\sigma_i^*$ is also (one of) the best response. Thus active contestant has no incentive to
deviate. This concludes the proof.

\end{proof}

\subsection{Monotonicity of Best Responses}\label{Prof.sa}
A critical property for the efficiency of our algorithms is the monotonicity of the best-response action share $\sigma_i$ with respect to the aggregate production $A$. The following is the proof of  Proposition \ref{sa}.
\begin{proof}
For the \emph{small elasticity} regime, the proof already exists in \cite{Cornes2005}.
For the case \emph{Medium and large elasticity}, by the \textit{first order condition},
$$
 b_2(A, \sigma_i; a_i, r_i)=a_ir_i^{r_i} (1 - \sigma_i)^{r_i} \sigma_i^{r_i - 1} - A = 0.
$$
Then
$$
\frac{d \sigma_i}{d A} = \frac{-1}{a_i r_i^{r_i} \left[ -r_i (1 - \sigma_i)^{r_i - 1} \sigma_i^{r_i - 1} + (r_i - 1) (1 - \sigma_i)^{r_i} \sigma_i^{r_i - 2} \right]} = \frac{1}{A \left(\frac{r_i - 1}{\sigma_i} - \frac{r_i}{1 - \sigma_i}\right)}
$$
So that $\frac{d \sigma_i}{d A} < 0$ when $\sigma_i > \frac{r_i-1}{2r_i-1}$. Since $\sigma \in[\frac{r_i-1}{r_i},1]$ and $\frac{r_i-1}{r_i} > \frac{r_i-1}{2r_i-1}$, then $\frac{d \sigma_i}{d A} < 0$.
And from \cite{Cornes2005}, when $A \in [\underline{A_i},\overline{A_i}]$, $\sigma^{BR}_i(A)$ might be $0$.
Hence we proof the case when $i \in \mathcal{I}^2$. This concludes the proof.
\end{proof}

\begin{figure}[htbp]
        \begin{subfigure}[b]{0.49\linewidth}
        \includegraphics[width=\linewidth]{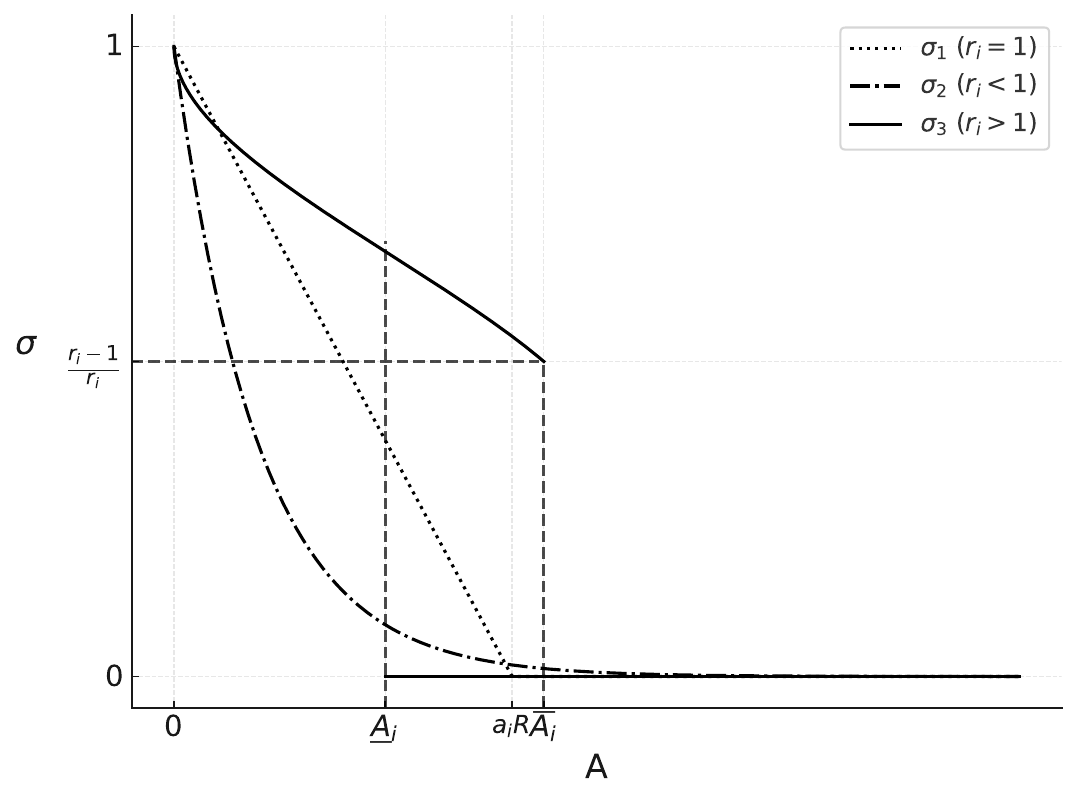}
        \caption{Best Response Action Share as a Function of Aggregate Production}
    \label{fig:as1}
    \end{subfigure}
    \hfill
    \begin{subfigure}[b]{0.48\linewidth}
        \includegraphics[width=\linewidth]{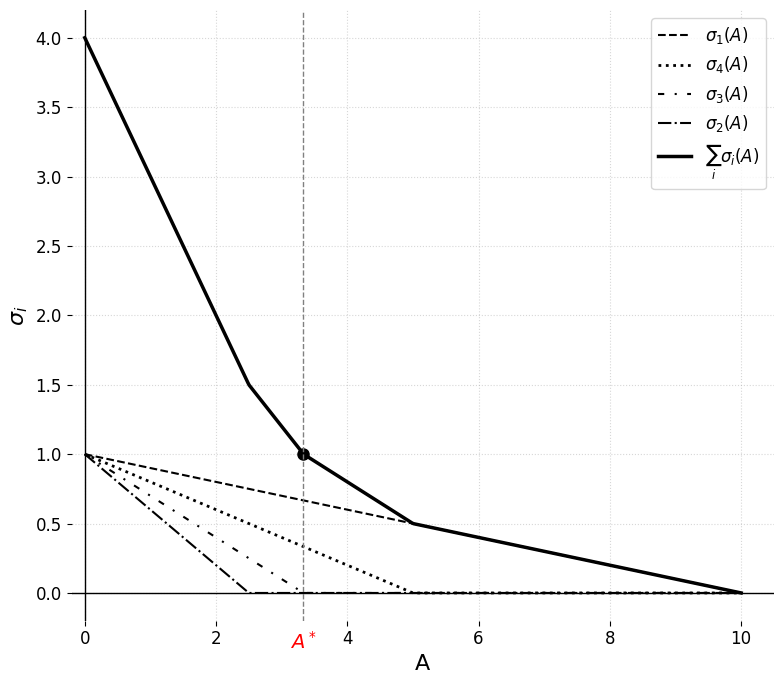}
        \caption{Best Response Action Share as a Function of Aggregate Production when $r_i = 1$.}
        \label{fig:as2}
    \end{subfigure}
\end{figure}

Figure~\ref{fig:as1} provides a clear visualization of how the best response action share $\sigma_i$ evolves with the aggregate production $A$ under different elasticity regimes ($r_i \leq 1, r_i = 1, r_i > 1$). The figure highlights the distinct patterns of behavior for contestants with varying elasticity parameters, demonstrating the critical role of $r_i$ in shaping strategic interactions. It shows that the best-response action share $ \sigma_i $ decreases with aggregate production $ A $ across all elasticity regimes, but the rate of decrease varies significantly. For $ r_i < 1 $, the decrease slows as $ A $ increases; for $ r_i = 1 $, the decrease is constant; and for $ r_i > 1 $, the decrease is non-monotonic, initially fast, then slowing, and accelerating again. Notably, in the range $ [\underline{A}, \bar{A}] $, $ \sigma_i $ can take two distinct values when $ r_i > 1 $, reflecting an ambiguity in participation behavior.

Figure~\ref{fig:as2} illustrates this relationship: as the aggregate production \( A \) increases, each contestant's action share \( \sigma_i \) decreases monotonically. This structural property enables the design of efficient algorithms that approximate the unique PNE.

\subsection{Impossibility Result for Convex Regime}\label{A}
Finally, we formally show why the purely convex regime ($r_i > 2$ for all players) is uninteresting from an equilibrium perspective (Proof of Proposition \ref{None}).
\begin{proof}
Consider a contestant $ i $ with $ r_i > 2 $. If contestant $ i $ is an active contestant in the equilibrium, then its action share $ \sigma_i $ must satisfy $ \sigma_i \geq \frac{r_i - 1}{r_i} $, as $ k(\bar{A_i};a_i) = \frac{r_i - 1}{r_i} $. Given that $ \frac{r_i - 1}{r_i} > \frac{1}{2} $ for $ r_i > 2 $, it follows that the action share $ \sigma_i $ for any active contestant $ i $ would exceed $ \frac{1}{2} $.

Since the sum of the action shares for all active contestants must equal $1$ in a PNE, it is impossible for more than one contestant to have an action share $ \sigma_i > \frac{1}{2} $. Because the number of the contestant $n \geq 2$, if a PNE exists, it must be the case that exactly one contestant is active, with this contestant holding an action share of $1$, while all other contestants have an action share of 0.

However, the optimal action share of $1$ leads to aggregation action $A = 0$. Since $\underline{A} > 0$, all other contestants also have an active action share of $1$ on $A = 0$.

Hence, a Tullock contest with $ r_i > 2 $ for all contestants admits no PNE.

\end{proof}

Also, we provide a simple example where rational solutions do not exist.
\paragraph{Example: Two-player contest with no rational solution.}\label{proof_easy_2}
A simple case when the PNE admits no rational solution.\\
Consider a Tullock contest with reward $R=1$ and two contestants: \\
$(a_1,r_1)=(1,0.5)$, $(a_2,r_2)=(1,1)$. $y_1=\sqrt{x_1}$,$y_2=x_2$. $A=y_1+y_2$. $\pi_1(y_1,y_2)=\frac{y_1}{A}-y_1^2$, $\pi_2(y_1,y_2)=\frac{y_2}{A}-y_2$.

\paragraph{First-order conditions.}
For an interior equilibrium, the first-order conditions are
\begin{align}
\frac{\partial \pi_1}{\partial y_1}
&=\frac{y_2}{(y_1+y_2)^2}-2y_1=0, \label{eq:foc1}\\
\frac{\partial \pi_2}{\partial y_2}
&=\frac{y_1}{(y_1+y_2)^2}-1=0. \label{eq:foc2}
\end{align}

From \eqref{eq:foc2}, we obtain $(y_1+y_2)^2=y_1$. Substituting into \eqref{eq:foc1} gives
$\frac{y_2}{y_1}=2y_1 \rightarrow y_2=2y_1^2$. Combining the two equations yields $2y_1^2=\sqrt{y_1}-y_1$.

Let $t=\sqrt{y_1}$. Then $y_1=t^2$ and $t$ satisfies the cubic equation $2t^3+t-1=0$.

By the rational root theorem, the only possible rational roots of
$2t^3+t-1=0$ are $\pm 1,\pm \tfrac12$, none of which satisfies the equation.
Hence the equation admits no rational solution.

\section{Efficient Algorithms for Restricted Regimes}\label{sec:eff_regime}

In this section, we provide the proofs and approximation guarantees for the regimes where the problem is computationally tractable: the small elasticity regime ($r_i \le 1$) and the logarithmic medium-elasticity regime ($m = O(\log n)$).

\subsection{Approximation Guarantees for Basic Algorithms in section \ref{easy_case}}\label{app:approx}
Before proving the main theorems, we establish the convergence rates for the Bisection Algorithm and Multi-Agent Mirror Descent (MAMD)) in the small elasticity regime.

\subsubsection{Bisection Algorithm}\label{bisection}
We prove that the bisection algorithm returns an $\varepsilon$-PNE in the \emph{small-elasticity regime}, i.e., when $r_i \le 1$ for every contestant $i$.

Let $A^*$ denote the unique equilibrium aggregate production, and let $\hat{A}$ be the aggregate production returned by the bisection algorithm. Each $\sigma_i$ is implicitly defined via the first-order condition $k_1(\sigma_i, A) = 0$ (see main text). The algorithm guarantees that the residual satisfies
\[
\left| \sum_{i=1}^n \hat{\sigma}_i - 1 \right| \le \delta,
\]
where each $\hat{\sigma}_i$ solves $k_1(\hat{\sigma}_i, \hat{A}) = 0$.

We show that this guarantees an $\varepsilon$-PNE when $\delta$ is chosen appropriately.

\paragraph{From Residual Error to Approximate Equilibrium}
\paragraph{Step 1: Bounding error in $\sigma_i$.}

Since each $\sigma_i$ is monotonic in $A$, we have
\[
\left| \hat{\sigma}_i - \sigma_i(A^*) \right|
\le \left| \sum_j \hat{\sigma}_j - 1 \right|
\le \delta.
\]

\paragraph{Step 2: Bounding utility error.}

Each contestant's utility is
\[
u_i = \sigma_i - x_i, \qquad\text{where } x_i = \left( \frac{\sigma_i A}{a_i} \right)^{1/r_i}.
\]

The first term changes by at most
\[
\left| \hat{\sigma}_i - \sigma_i(A^*) \right| = |\hat{\sigma}_i - \sigma_i(A^*)| \le \delta.
\]

For the second term, let
\[
X := \frac{\sigma_i(A^*)\,A^*}{a_i}, \qquad \hat{X} := \frac{\hat{\sigma}_i\,\hat{A}}{a_i},
\]
so that
\[
|x_i(\hat{A}) - x_i(A^*)| = \left| \hat{X}^{1/r_i} - X^{1/r_i} \right|.
\]

Now note that the map $t \mapsto t^{1/r_i}$ is Lipschitz continuous on the interval $[X_{\min}, X_{\max}]$ for all $i$, with Lipschitz constant
\[
L' := \max_{i} \frac{1}{r_i} \left( \frac{\bar{a} A_{\max}}{\underline{a}} \right)^{\frac{1 - r_i}{r_i}} \le \frac{1}{r_{\min}} \left( \frac{\bar{a} A_{\max}}{\underline{a}} \right)^{\frac{1 - r_{\min}}{r_{\min}}}.
\]

Next, we bound $|\hat{X} - X|$ as
\[
|\hat{X} - X|
= \frac{1}{a_i} \left| \hat{\sigma}_i \hat{A} - \sigma_i(A^*) A^* \right|
\le \frac{1}{\underline{a}} \left( A_{\max}\,\delta + \sigma_{\max}\,\delta \right)=\frac{\left( A_{\max}\,\delta + \sigma_{\max}\,\delta \right)}{\underline{a}} 
= L''\,\delta,
\]
where $\sigma_{\max} \le 1$ and $L'' := \frac{A_{\max} + 1}{\underline{a}}$.

Putting these together:
\[
|x_i(\hat{A}) - x_i(A^*)| \le L' L'' \delta.
\]

\paragraph{Step 3: Combining both.}

The total utility difference satisfies
\[
|u_i(\hat{A}) - u_i(A^*)|
\le \delta + L' L''\,\delta
= (1 + L' L'')\,\delta.
\]

Since the action shares of all active contestants are computed up to precision $\delta$,
the deviation in the aggregate constraint satisfies
$\bigl|\sum_i \sigma_i - 1\bigr| \le n\delta$.
By the Lipschitz bounds established above, this induces a utility deviation of at most
$(1+L' L'')\,n\delta$.
Hence, choosing
\[
\delta = \frac{\varepsilon}{n(1+L' L'')}
\]
guarantees that the bisection algorithm returns an $\varepsilon$-PNE.

\subsubsection{Multi-Agent Mirror Descent}

We now show that the MAMD algorithm computes an $\varepsilon$-PNE in the small elasticity regime. Our argument proceeds in two steps: we first justify that without loss of generality, we may exclude the Tullock instances with a trivial PNE $\boldsymbol{x} = \boldsymbol{0}$ from consideration; then we prove that all limit points returned by MAMD are $\varepsilon$-approximate equilibria.

\paragraph{Step 1: Excluding Tullock instances with zero PNEs.}
Observe that $\boldsymbol{x} = \boldsymbol{0}$ is not a valid equilibrium under the small elasticity regime ($r_i \le 1$), unless each contestant’s marginal utility at zero effort is finite and nonpositive. However, in the Tullock setting, contestant $i$'s marginal utility when others exert zero effort is
\[
\left.\frac{\partial u_i}{\partial x_i}\right|_{\boldsymbol{x}_{-i}=0}
= \left.\frac{a_i r_i x_i^{r_i - 1}}{a_i x_i^{r_i}} - 1\right|_{x_i \to 0}
= \infty.
\]
Hence, $\boldsymbol{x} = \boldsymbol{0}$ cannot be a Nash equilibrium unless the cost function has infinite slope at $x_i = 0$, which does not hold in our model. We may therefore assume that the problem instance we consider does not contain a PNE at $\bm{0}$. 

Let $\boldsymbol{x}^*$ be a PNE and $j=\arg\max_{i\in [n]}x_i^*$. Suppose $x_k^*$ is the second largest element of $\boldsymbol{x}^*$ except $x_j^*$. Now we claim that there exists $\delta>0$ such that one of the following claims must hold: 1. $x_j^*\geq x_k^*\geq \delta$, 2. $\arg\max_{x'} u_j(x', \boldsymbol{x}_{-j}^*)\geq \delta$. This is because if condition 1 does not hold, we have $\boldsymbol{x}_{-j}^*=\bm{0}_{n-1}$. However, since $\bm{0}_n$ is not a PNE, the best response for contestant $j$ must not be zero given $\boldsymbol{x}_{-j}^*=\bm{0}_{n-1}$, meaning $\arg\max_{x'} u_j(x', \boldsymbol{x}_{-j}^*)>0$. As a result, given any instance such that $\bm{0}$ is not a PNE, we can find such $\delta>0$. In the following, we will use the property of $\delta$ to complete the proof.

\paragraph{Step 2: Approximate optimality of limit point.}
Since MAMD is a convergent no-regret procedure under smooth, strictly monotone games (as ensured by the socially concave structure of the small elasticity regime), the iterate $\hat{\boldsymbol{x}}$ returned by MAMD lies in a small neighborhood around an exact PNE $\boldsymbol{x}^*$. Therefore, we may assume $\|\boldsymbol{x^*}-\hat{\boldsymbol{x}}\|\leq \frac{\delta_1}{2}$ for some $0<\delta_1<\delta$.

Next, we show that $\hat{\boldsymbol{x}}$ is an $\varepsilon$-PNE, i.e., there exists $\varepsilon>0$ such that for any $i$,

\begin{equation}\label{eq:120}
u_i(\hat{x}_i; \hat{\boldsymbol{x}}_{-i}) \ge \max_{x_i' \in \mathbb{R}_{\ge 0}} u_i(x_i'; \hat{\boldsymbol{x}}_{-i}) - \varepsilon.    
\end{equation}

Since $\boldsymbol{x}^*$ is a PNE, by definition we have 
\begin{equation}\label{eq:126}
u_i(x^*_i; \boldsymbol{x}^*_{-i}) \ge \max_{x_i' \in \mathbb{R}_{\ge 0}} u_i(x_i'; \boldsymbol{x}^*_{-i}).
\end{equation}

To show that Eq. \eqref{eq:120} holds, it is sufficient to prove Lipschitz continuity condition for each $u_i$, as such a property implies that $|u_i(\hat{x}_i; \hat{\boldsymbol{x}}_{-i})-u_i(x^*_i; \boldsymbol{x}^*_{-i})|$ and $|u_i(x_i'; \hat{\boldsymbol{x}}_{-i})-u_i(x_i'; \boldsymbol{x}^*_{-i})|$ are both small when $\boldsymbol{x}^*-\hat{\boldsymbol{x}}$ are sufficiently close. To show this, observe that each $u_i(\boldsymbol{x})$ is differentiable and concave in $\boldsymbol{x}$ and therefore it holds that for any $\boldsymbol{x},\boldsymbol{x'}$,

\[
|u_i(\boldsymbol{x})-u_i(\boldsymbol{x'})|\leq |\nabla u_i(\boldsymbol{x})\cdot(\boldsymbol{x}-\boldsymbol{x'})|\leq \|\nabla u_i(\boldsymbol{x})\|\cdot \|\boldsymbol{x}-\boldsymbol{x'}\|.
\]

As a result, 
\begin{equation}\label{eq:159}
|u_i(\hat{x}_i; \hat{\boldsymbol{x}}_{-i})-u_i(x^*_i; \boldsymbol{x}^*_{-i})| \leq \|\nabla u_i(\boldsymbol{x^*})\|\cdot \|\boldsymbol{x^*}-\hat{\boldsymbol{x}}\|\leq C_1\cdot \frac{\delta_1}{2}.    
\end{equation}

Next we show that for any $x'_i$ and $x''_i=\arg\max_{x_i\in \mathbb{R}_{\geq 0}} u_i(x_i,\boldsymbol{x^*}_{-i})$, we have 
\begin{equation}\label{eq:162}
|u_i(x'_i; \hat{\boldsymbol{x}}_{-i})-u_i(x''_i; \boldsymbol{x}^*_{-i})| \leq \|\nabla u_i(x''_i,\boldsymbol{x^*}_{-i})\|\cdot \|\boldsymbol{x^*}_{-i}-\hat{\boldsymbol{x}}_{-i}\|\leq C_1\cdot \frac{\delta_1}{2}.    
\end{equation}

\begin{enumerate}
    \item if $j\neq i$, we have $\|(x''_i,\boldsymbol{x^*}_{-i})\|_{\infty}\geq x_j^*\geq\delta$ and, therefore, by Lemma \ref{lm:160}, $\|\nabla u_i(x''_i,\boldsymbol{x^*}_{-i})\|\leq C_1$.
    \item if $j=i$, according to the definition of $\delta$, there are two cases:
    \begin{enumerate}
        \item if $x_j^*\geq x_k^*\geq \delta$, it holds that $\|(x''_i,\boldsymbol{x^*}_{-i})\|_{\infty}\geq x_k^* \geq \delta$. By Lemma \ref{lm:160}, $\|\nabla u_i(x''_i,\boldsymbol{x^*}_{-i})\|\leq C_1$.
        \item if $\arg\max_{x'} u_j(x', \boldsymbol{x}_{-j}^*)\geq \delta$, we have $\|(x''_i,\boldsymbol{x^*}_{-i})\|_{\infty}\geq x''_i \geq \delta$. By Lemma \ref{lm:160}, $\|\nabla u_i(x''_i,\boldsymbol{x^*}_{-i})\|\leq C_1$.

    \end{enumerate}
\end{enumerate}


Combining Eq. \eqref{eq:126}, \eqref{eq:159}, \eqref{eq:162}, we conclude that Eq. \eqref{eq:120} holds with $\varepsilon=C_1\cdot \delta_1$. Because $\delta_1>0$ can be arbitrarily small as MAMD algorithm iterates, $\varepsilon \rightarrow 0$ and $\hat{\boldsymbol{x}}$ is indeed an $\varepsilon$-NE.

\begin{lemma}\label{lm:160}
For any $\boldsymbol{x},\delta>0$ such that $\|\boldsymbol{x}\|_{\infty}\geq\delta$, there exists a positive constant $C_1$ such that $\|\nabla u_i(\boldsymbol{x})\|_2\leq C_1,\forall i\in[n]$.
\end{lemma}

\begin{proof}[Proof of Lemma \ref{lm:160}]
Fix any \(\delta>0\) and consider any profile \(\boldsymbol{x}\in\mathbb{R}_{\ge0}^n\) with \(\|\boldsymbol{x}\|_\infty\ge\delta\).  Let
\[
F(\boldsymbol{x})
=\sum_{j=1}^n f_j(x_j)
\;=\;
\sum_{j=1}^n a_j\,x_j^{r_j},
\]
and write \(a_{\min}=\min_j a_j\), \(a_{\max}=\max_j a_j\), \(r_{\min}=\min_j r_j\), \(r_{\max}=\max_j r_j\).  Since \(x_j\ge0\) and \(\|\boldsymbol x\|_\infty\ge\delta\), we have
\[
F(\boldsymbol{x})
\;\ge\;
a_{\min}\,\delta^{\,r_{\min}}
\quad\Longrightarrow\quad
F(\boldsymbol{x})^2 \;\ge\; a_{\min}^2\,\delta^{2r_{\min}}.
\]

The utility is
\[
u_i(\boldsymbol x)
=\frac{f_i(x_i)}{F(\boldsymbol x)} \;-\; x_i,
\]
so for \(k\neq i\),
\[
\frac{\partial u_i}{\partial x_k}
=-\frac{f_i(x_i)\,f_k'(x_k)}{F(\boldsymbol x)^2}
\quad\Longrightarrow\quad
\Bigl|\partial_{x_k}u_i(\boldsymbol x)\Bigr|
\;\le\;
\frac{a_{\max}\,\|\boldsymbol x\|_\infty^{r_i}\;\cdot\;a_{\max}\,r_{\max}\,\|\boldsymbol x\|_\infty^{r_{\max}-1}}
            {a_{\min}^2\,\delta^{2r_{\min}}}
=:B_1.
\]
For \(k=i\),
\[
\frac{\partial u_i}{\partial x_i}
=\frac{f_i'(x_i)\,F(\boldsymbol x)-f_i(x_i)\,f_i'(x_i)}{F(\boldsymbol x)^2}-1,
\]
hence
\[
\Bigl|\partial_{x_i}u_i(\boldsymbol x)\Bigr|
\;\le\;
\frac{f_i'(x_i)}{F(\boldsymbol x)} \;+\;1
\;\le\;
\frac{a_{\max}\,r_{\max}\,\|\boldsymbol x\|_\infty^{\,r_{\max}-1}}
         {a_{\min}\,\delta^{\,r_{\min}}}
\;+\;1
=:B_2.
\]

Putting these together, the Euclidean norm of the gradient satisfies
\[
\bigl\|\nabla u_i(\boldsymbol x)\bigr\|_2
\;\le\;
\sqrt{\sum_{k=1}^n \bigl|\partial_{x_k}u_i(\boldsymbol x)\bigr|^2}
\;\le\;
\sqrt{n}\,\max\{B_1,B_2\}.
\]
Setting
\[
C_1 \;=\;\sqrt{n}\,\max\{B_1,B_2\}
\]
yields the desired uniform bound
\(\|\nabla u_i(\boldsymbol x)\|_2\le C_1\) for all \(i\) and all \(\boldsymbol x\) with \(\|\boldsymbol x\|_\infty\ge\delta\).
\end{proof}

\subsection{Proof of Efficiency in Small Elasticity Regime}\label{tthe2}
Building on the bisection guarantee, we prove Theorem \ref{the2}.

\begin{proof}

We prove that when at most $O(\log n)$ contestants have medium elasticity
$r_i\in(1,2]$, the existence of PNE can be decided in
polynomial time.

By Proposition~\ref{PNEProp}, PNE is fully characterized by a triple
$(A^*,\boldsymbol{\sigma}^*,\mathcal{I}^A)$ consisting of the aggregate production,
the action-share profile, and the set of active contestants. The algorithm searches
for such a triple by enumerating candidate active sets and, for each candidate,
verifying whether a feasible aggregate production $A^*$ exists.

Let $\mathcal{I}^{M}=\{i:r_i\in(1,2]\}$ denote the set of contestants with medium
elasticity, with $|\mathcal{I}^{M}|=m$, and let
$\mathcal{I}^{L}=\{i:r_i>2\}$ denote the set of contestants with large elasticity.
Contestants with $r_i\le 1$ are always active and hence need not be enumerated.
By Corollary~\ref{act}, at most one contestant in $\mathcal{I}^{L}$ can be active
in any PNE.

Accordingly, the algorithm enumerates all $(|\mathcal{I}^{L}|+1)$ possibilities
for the active contestant in $\mathcal{I}^{L}$ (including the case where none is
active), and all $2^m$ subsets of $\mathcal{I}^{M}$. Thus, the total number of
candidate active sets is at most
\[
(|\mathcal{I}^{L}|+1)\cdot 2^m \;\le\; (n+1)\cdot 2^m  = O(n^{c+1}) = n^{O(1)}.
\]
Since $m=O(\log n)$, this quantity is polynomial in $n$.

For a fixed candidate active set $\mathcal{I}^A$, the equilibrium conditions in
Proposition~\ref{PNEProp} impose bounds on the aggregate production $A$.
Specifically, for each active contestant $i\in\mathcal{I}^A$ with $r_i>1$ we must have
$A\le \bar A_i$, while for each inactive contestant $j\notin\mathcal{I}^A$ with $r_j>1$
we must have $A\ge \underline A_j$. Hence the feasible range of $A$ is
\[
A \in
\Bigl[
\max_{j\notin\mathcal{I}^A}\underline A_j,
\;
\min_{i\in\mathcal{I}^A}\bar A_i
\Bigr].
\]
If this interval is empty, the candidate active set cannot support a PNE and is
discarded.

\paragraph{Complexity Analysis.}
When the feasible interval is nonempty, the algorithm performs a binary search
over $A$ within this interval to determine whether there exists an $A^*$ such that
the induced best-response action shares satisfy
$\sum_{i\in\mathcal{I}^A}\sigma_i(A)=1$.

The number of bisection steps is determined by the target accuracy in $A$.
By Appendix~\label{bisection}, to guarantee an $\varepsilon$-accurate solution it suffices to search up to
precision $\delta$ in $A$, where $\delta = \Theta(\varepsilon/nL)$ and $L$ is the relevant Lipschitz
constant. Hence the bisection requires
\[
O\!\left(\log\frac{\bar A}{\delta}\right)
=O\!\left(\log\frac{\bar A nL}{\varepsilon}\right)
=O\!\left(\log\frac{\bar An}{\varepsilon}\right)
\]
iterations (absorbing constant factors into the $O(\cdot)$ notation).

At each bisection query $A$, we compute every contestant's best-response action share
$\sigma_i(A)$. When $r_i\neq 1$, $\sigma_i(A)$ has no closed form and is obtained by solving the
first-order condition using a numerical method (e.g., Newton's method) to accuracy $\delta'$.
The solver converges in at most $O(\log(1/\delta'))$ iterations; setting $\delta'=\Theta(\varepsilon/c)$
for an appropriate constant $c$ yields $O(\log(c/\varepsilon))=O(\log(1/\varepsilon))$ iterations per
contestant. Therefore, evaluating all $n$ best responses at a given $A$ costs
$O\!\left(n\log\frac{1}{\varepsilon}\right)$ time (up to constant-time arithmetic per iteration).

Putting the above components together, the total running time of the algorithm is
\[
n^{O(1)}
\;\cdot\;
O\!\left(\log\frac{\bar An}{\varepsilon}\right)
\;\cdot\;
O\!\left(n\log\frac{1}{\varepsilon}\right),
\]
which is polynomial in $n$ and $1/\varepsilon$ since $m=O(\log n)$.
Hence, when at most $O(\log n)$ contestants have medium elasticity, the existence
of a PNE can be decided in polynomial time. If such an equilibrium exists, the
algorithm explicitly identifies a corresponding $\varepsilon$-PNE.

\end{proof}

\section{Computational Hardness Proofs}\label{sec:hard_proofs}

In this section, we provide the complete reduction chain establishing that determining the existence of a PNE is NP-complete. Our proof strategy relies on transforming the combinatorial difficulty of the \textit{Subset Sum} problem into the strategic ambiguity of a Tullock contest. The reduction proceeds in three logical steps:
\begin{enumerate}
    \item \textbf{Variant Definition:} We first introduce a specific variant of the Subset Sum problem, which we term \textit{Subset Sum with Large Targets} (\texttt{SSLT}), and prove its hardness. This variant is necessary because mapping discrete numbers to player participation constraints requires strictly positive, bounded targets.
    \item \textbf{Core Reduction:} We then construct a specific "All-Medium" Tullock contest instance that mimics the structure of \texttt{SSLT}. This step proves that the problem is NP-complete when $m=n$.
    \item \textbf{Generalization via Padding:} Finally, we extend this hardness to the general regime where medium-elasticity players are sparse ($m \approx \log n$) by padding the hard instance with dummy players.
\end{enumerate}

\subsection{Hardness of Subset Sum with Large Targets}\label{C}
Standard Subset Sum instances may involve targets that are small or zero, which are difficult to map directly to the strictly positive equilibrium aggregates in Tullock contests. To bridge this gap, we define \texttt{SSLT}, where the target sum is sufficiently large to force the inclusion of specific "heavy" elements. Then we show that \textit{SSLT} is also hard by the proof of Lemma \ref{ssbm}.

\begin{proof}
We prove that the Subset Sum with Large Targets (\texttt{SSLT}) problem is NP-complete by showing that (i) it is NP-hard via a reduction from \textit{Subset Sum}, and (ii) it belongs to NP.

\paragraph{NP-hardness.} Given an instance of \textit{Subset Sum} with a set $ Z = \{z_1, z_2, \dots, z_n\} $ and a target sum $ T $, we construct an instance of \texttt{SSLT} by introducing two additional elements $S$, where $S = \sum_{z \in Z} z$. And we define the target sum as $T' = T + 2S$.

Since $ T' > 2S $, any valid subset must include both of $ S $. The remaining elements must then sum to $ T $, ensuring a solution to the original \textit{Subset Sum} instance.

Conversely, if a solution exists for \textit{Subset Sum}, selecting the corresponding subset from $ Z $ along with both $ S $ values achieves $ T' $. Since this reduction runs in polynomial time and \textit{Subset Sum} is NP-complete, it follows that \texttt{SSLT} is NP-hard.

\paragraph{Membership in NP} Given a subset of numbers from the input set, we can verify in polynomial time whether their sum equals $ T' $. This confirms that \texttt{SSLT} belongs to NP.

Since \texttt{SSLT} is both NP-hard and in NP, it is NP-complete.
\end{proof}

\subsection{Reduction to All-Medium Contest}\label{D}
With the \texttt{SSLT} hardness established, we now translate this combinatorial problem into a game-theoretic one. The key insight is to design contestant parameters $(a_i, r_i)$ such that each contestant effectively faces a binary choice: either participate with a specific effort level (corresponding to "selecting" a number) or stay inactive (corresponding to "not selecting"). The following is the proof of Lemma \ref{npc}

\begin{proof}
Consider an instance of \texttt{SSLT} problem defined by a set of positive integers $ Z = \{z_1, \dots, z_n\} $ where $ \max_{z \in \textbf{Z}} z \leq \frac{1}{2} \bar{z} $, and a target sum $ \bar{z} $. Let $ z' = \min_{i \in \{1,\dots,n\}} z_i $. We construct a corresponding Tullock contest with $ n+1 $ contestants, where the contest parameters are defined as follows:
    $$
    \begin{cases}
         r_i = \frac{1}{1 - \frac{z_i}{\bar{z}}} & \text{for } i \in \{1, \dots, n\}\\
         r_i = \frac{1}{\frac{z'}{\bar{z}}}  & \text{for } i = n + 1
    \end{cases}
    $$
    Given that $ \max_{z \in \textbf{Z}} z \leq \frac{1}{2} \bar{z} $, it follows that $ r_i \in (1, 2] $ for all $ i \in \{1, \dots, n+1\} $.
    
    The corresponding production function parameters are defined as:
    $$
    a_i = \begin{cases}
        \frac{\bar{z}}{(\frac{r_i-1}{r_i})^{r_i-1}} & \text{for } i \in \{1, \dots, n\}\\
        \frac{\bar{z}}{ \frac{(r_{i}-1)^{r_{i}-1}}{{r_i}^{r_i}}} & \text{for } i = n+1
    \end{cases}
    $$
    
    We begin by considering the constructed Tullock contest, where the goal is to show that this contest corresponds to an instance of the \texttt{SSLT}. Specifically, we aim to prove that there is a PNE in the Tullock contest if and only if the \texttt{SSLT} has a solution.
    
    First, consider the case where the \texttt{SSLT} problem has no solution. In the corresponding Tullock contest, let us examine the instance we have constructed.
    $$
    \begin{cases}
    \bar{A_i} = (\frac{r_i-1}{r_i})^{r_i-1} \cdot a_i = \bar{z} & \text{for } i \in \{1, \dots, n\}\\
    \underline{A}_{n+1} = \frac{(r_i-1)^{r_i-1}}{{r_i}^{r_i}} \cdot a_i = \bar{z} & \text{for } i = n + 1
    \end{cases}
    $$
Let $ A^* $ denote the aggregate production of all contestants. We first establish that if $ A^* \neq \bar{z} $, a PNE cannot exist.

\paragraph{Case 1: $ A^* < \bar{z} $.}  
In this case, the action share of contestant $ n+1 $ is given by $ \sigma_{n+1} = k(A^*; a_{n+1}, r_{n+1}) $. Since $ k(A; a_i, r_i) $ is decreasing over $ [0, \bar{A}_{n+1}] $, we have:
\[
k(A^*; a_{n+1}, r_{n+1}) > k(\bar{A}_{n+1}; a_{n+1}, r_{n+1}) = \frac{r_{n+1} - 1}{r_{n+1}} = 1 - \frac{z'}{\bar{z}}.
\]
Thus, $ \sigma_{n+1} \in (1 - \frac{z'}{\bar{z}}, 1) $.

For the other contestants $ i \in \{1, \dots, n\} $, each action share satisfies $ \sigma_i \in \{k(A^*; a_i, r_i), 0\} $. Since $ k(A^*; a_i, r_i) > k(\bar{A}_i; a_i, r_i) = \frac{z_i}{\bar{z}} \geq \frac{z'}{\bar{z}} $, the sum of all action shares either exceeds 1 or equals $ k(A^*; a_{n+1}, r_{n+1}) $, but never exactly 1. Thus, no equilibrium exists when $ A^* < \bar{z} $.

\paragraph{Case 2: $ A^* > \bar{z} $.}  
Here, we have $ \sigma_i = 0 $ for all $ i \in \{1, \dots, n\} $, meaning only contestant $ n+1 $ could be active. Since this results in $ \sum_{i \in \mathcal{I}^A} \sigma_i < 1 $, no equilibrium is possible in this case.

\paragraph{Case 3: $ A^* = \bar{z} $.}  
By the above arguments, a necessary condition for equilibrium is $ A^* = \bar{z} $. Next, we establish that contestant $ n+1 $ cannot be active in equilibrium. Given the construction, contestant $ n+1 $'s optimal action share satisfies:
\[
\sigma_{n+1} > 1 - \min_{i \in \{1, \dots, n\}} \sigma_i.
\]
If contestant $ n+1 $ were active, their action share would exceed the feasible limit, violating the equilibrium condition $ \sum_{i \in \mathcal{I}^A} \sigma_i = 1 $. Thus, contestant $ n+1 $ must be inactive.

\paragraph{ Connecting to Subset Sum} With contestant $ n+1 $ excluded, we must determine whether the remaining $ n $ contestants can form an equilibrium with $ A = \bar{z} $. Each contestant $ i $ has the option to set $ \sigma_i = k(A^*; a_i,r_i) $ or $ \sigma_i = 0 $, corresponding to including or excluding $ z_i $ in a subset sum.

Thus, finding an equilibrium is equivalent to computing the Subset Sum problem: determining whether a subset $ S \subseteq Z $ exists such that:
\[
\sum_{z \in S} z = \bar{z}.
\]

Therefore, we have shown that \texttt{SSLT} has a solution if and only if the corresponding Tullock contest instance has a PNE, with the equilibrium directly corresponding to the solution of the \texttt{SSLT}.
\end{proof}

\subsection{Inapproximability Results}\label{sec:proof_inapprox}

In this section, we provide the detailed proofs for the inapproximability lower bounds discussed in Section \ref{sec:hard_proofs}. We explicitly derive the exponential gap arising from the discrete nature of the problem input and demonstrate why this precludes approximation algorithms with logarithmic dependence on precision.

\subsubsection{Proof of Lemma \ref{lem:gap}}

\begin{proof}
    Recall the reduction construction from Lemma \ref{npc}. We constructed a Tullock contest instance such that at the critical aggregate production level $A = \bar{z}$, the best-response action share for any contestant $i$ is effectively binary:
    \[
    \sigma_i(A) \in \left\{ \frac{z_i}{\bar{z}}, \, 0 \right\}.
    \]
    Let $\mathcal{I}^A$ denote the set of active contestants in a given strategy profile. The sum of action shares is given by:
    \[
    \sum_{i \in \mathcal{I}^A} \sigma_i = \sum_{i \in \mathcal{I}^A} \frac{z_i}{\bar{z}} = \frac{1}{\bar{z}} \left( \sum_{i \in \mathcal{I}^A} z_i \right).
    \]
    By hypothesis, the instance $I$ does not admit a PNE. This implies that the underlying \texttt{SSLT} instance is a "No" instance. Consequently, for any subset of contestants $\mathcal{I}^A$, the subset sum of values does not equal the target:
    \[
    \sum_{i \in \mathcal{I}^A} z_i \neq \bar{z}.
    \]
    
    We now quantify the minimum non-zero difference $|\sum_{i \in \mathcal{I}^A} z_i - \bar{z}|$.
    
    Since the problem input is encoded in binary, the values $\{z_1, \dots, z_n, \bar{z}\}$ are rational numbers. Let the total bit length of the input be $L$. Each number $z_i$ (and $\bar{z}$) can be represented as a fraction $\frac{p_i}{q_i}$, where $p_i, q_i$ are integers encoded within the input size.
    
    Let $D$ be the common denominator of the set $\{z_1, \dots, z_n, \bar{z}\}$. A valid choice for $D$ is the product of all individual denominators:
    \[
    D = \left(\prod_{i=1}^n q_i\right) \cdot q_{\bar{z}}.
    \]
    Using this common denominator, we can rewrite each value as an integer scaled by $1/D$. Specifically, let $z_i = \frac{k_i}{D}$ and $\bar{z} = \frac{K}{D}$, where $k_i$ and $K$ are integers.
    
    Substituting these into the subset sum difference:
    \[
    \left| \sum_{i \in \mathcal{I}^A} z_i - \bar{z} \right| = \left| \sum_{i \in \mathcal{I}^A} \frac{k_i}{D} - \frac{K}{D} \right| = \frac{1}{D} \underbrace{\left| \sum_{i \in \mathcal{I}^A} k_i - K \right|}_{\text{Integer Difference}}.
    \]
    Since $\sum z_i \neq \bar{z}$, it follows that $\sum k_i \neq K$. Because $k_i$ and $K$ are integers, the absolute difference $|\sum k_i - K|$ must be at least 1. Therefore:
    \[
    \left| \sum_{i \in \mathcal{I}^A} z_i - \bar{z} \right| \ge \frac{1}{D}.
    \]
    
    Now, we relate this back to the market-clearing condition. The deviation is:
    \[
    \left| \sum_{i \in \mathcal{I}^A} \sigma_i - 1 \right| = \left| \frac{1}{\bar{z}} \sum_{i \in \mathcal{I}^A} z_i - 1 \right| = \frac{1}{\bar{z}} \left| \sum_{i \in \mathcal{I}^A} z_i - \bar{z} \right| \ge \frac{1}{\bar{z}} \cdot \frac{1}{D} = \frac{1}{\bar{z} D}.
    \]
    
    Finally, we bound the magnitude of $\frac{1}{\bar{z} D}$ in terms of the input size $n$ (or total bit length $L$). In standard binary encoding, the value of an integer (or denominator) encoded with $b$ bits is at most $2^b$. Since the total input size is $L$ (which is polynomial in $n$), the product of denominators $D$ and the value $\bar{z}$ are bounded by $2^{poly(L)}$. Specifically, $\bar{z} D \le 2^{c \cdot L}$ for some constant $c$.
    
    Defining $\Delta_{\min} = \frac{1}{\bar{z} D}$, we conclude:
    \[
    \left| \sum_{i \in \mathcal{I}^A} \sigma_i - 1 \right| \ge \Delta_{\min} \ge \frac{1}{2^{poly(L)}} = 2^{-poly(n)}.
    \]
    This completes the proof.
\end{proof}

\subsubsection{Proof of Theorem \ref{thm:inapprox}}

\begin{proof}
    We proceed by contradiction. Assume there exists an algorithm $\mathcal{A}_{fast}$ that computes an $\varepsilon$-PNE for the general Tullock contest in time $T(n, \varepsilon) = poly(n, \log(1/\varepsilon))$.
    
    We construct a specific instance to solve the \texttt{SSLT} problem, which is NP-complete. Let an instance of \texttt{SSLT} be given by inputs $Z$ and target $\bar{z}$. We perform the reduction described in Lemma \ref{npc} to create a corresponding Tullock contest instance.
    
    According to Lemma \ref{lem:gap}, this constructed instance exhibits a discrete gap: if no PNE exists, any candidate strategy profile violates the market-clearing condition by at least $\Delta_{\min}$. We choose the precision parameter $\varepsilon$ to be strictly smaller than half of this gap:
    \[
    \varepsilon := \frac{1}{2} \Delta_{\min} = \frac{1}{2\bar{z}D}.
    \]
    Crucially, because $\Delta_{\min} = 2^{-poly(n)}$, the term $\log(1/\varepsilon)$ is polynomial in the input size $n$:
    \[
    \log(1/\varepsilon) = \log(2\bar{z}D) \approx O(poly(n)).
    \]
    
    Now, we run the hypothetical algorithm $\mathcal{A}_{fast}$ on this instance with the chosen $\varepsilon$. The runtime would be:
    \[
    T(n, \varepsilon) = poly(n, \log(1/\varepsilon)) = poly(n, poly(n)) = poly(n).
    \]
    Thus, $\mathcal{A}_{fast}$ would terminate in polynomial time.
    
    Let the algorithm return a solution with market-clearing error $\delta$, where $\delta \le \varepsilon$. We use this output to decide the \texttt{SSLT} problem:
    \begin{itemize}
        \item If the \texttt{SSLT} instance has a solution ("Yes" instance), a PNE exists with error $0$. The algorithm will return a solution within error $\varepsilon$.
        \item If the \texttt{SSLT} instance has no solution ("No" instance), by Lemma \ref{lem:gap}, any possible strategy profile has error at least $\Delta_{\min}$. Since $\varepsilon < \Delta_{\min}$, it is impossible for the algorithm to find a solution with error $\le \varepsilon$ (assuming the algorithm is correct for $\varepsilon$-PNE). Or more formally, if the algorithm returns a state with error $\le \varepsilon < \Delta_{\min}$, it \textit{must} be that a true PNE exists (or the algorithm implies the gap is violated, which contradicts the "No" instance property).
    \end{itemize}
    
    Therefore, the existence of $\mathcal{A}_{fast}$ would allow us to distinguish between "Yes" and "No" instances of the NP-complete \texttt{SSLT} problem in polynomial time. This implies $P = NP$. By contradiction, no such algorithm exists.
\end{proof}

\section{FPTAS Analysis and Correctness}\label{sec:fptas_proofs}

This section provides the theoretical guarantees for the Fully Polynomial-Time Approximation Scheme (FPTAS). Our algorithm circumvents the NP-hardness by relaxing the exact equilibrium condition. The correctness of this approach relies on two pillars:
\begin{enumerate}
    \item \textbf{Discretization Validity:} We must show that searching over a polynomial-size grid is sufficient to find a candidate solution if one exists.
    \item \textbf{Stability of Approximation:} We must prove that a solution which satisfies the market-clearing condition \textit{approximately} corresponds to a valid $\varepsilon$-PNE in terms of \textit{utility}.
\end{enumerate}

\subsection{Discussion on Complexity: FPTAS vs. PPTAS}
\label{sec:pptas_discussion}

In the main text (Section \ref{sec:approx}), we established the efficiency of our algorithm under the condition that the elasticity parameters for medium contestants are strictly bounded away from unity. Specifically, for all $i$ such that $r_i \in (1, 2]$, we assumed $r_i \ge 1 + \delta$ for some constant $\delta > 0$. In this section, we provide the theoretical justification for this condition by analyzing the Lipschitz constant $L_u$ governing the convergence of our approximation.

As derived in the proof of Theorem \ref{thm:fptas_main} and the derivation of the Lipschitz constant $L_u$ (see Eq. (\ref{lipschize})), the conversion factor $L_u$ between the algorithmic error and the utility deviation contains a critical dependency on the minimum elasticity among medium contestants, denoted as $r'_{\min} = \min \{r_i : r_i \in (1, 2]\}$. Explicit inspection of $L_u$ reveals:
\[
L_u \propto \frac{1}{r'_{\min} - 1}.
\]

\paragraph{Justification for FPTAS With assumption.}
To achieve a \textit{Fully Polynomial-Time Approximation Scheme} (FPTAS), the runtime must be polynomial in the input size $n$ and $1/\varepsilon$, and independent of the specific numerical values of the input parameters. By imposing the assumption (i.e., $r_i \ge 1 + \delta$ for some constant $\delta > 0$), we ensure that:
\[
\frac{1}{r_i - 1} \le \frac{1}{\delta} = O(1).
\]
Consequently, the Lipschitz constant $L_u$ becomes a bounded value independent of the input parameters' proximity to $1$. This guarantees that the grid size required by Algorithm \ref{alg:FPTAS} grows only polynomially with $n$ and $1/\varepsilon$, satisfying the strict definition of an FPTAS.

\paragraph{Pseudo-Polynomial Time Without assumption.}
If we relax the assumption and allow medium-elasticity parameters to be arbitrarily close to $1$, the term $\frac{1}{r_i - 1}$ is no longer bounded by a constant. In this general case, the runtime of our algorithm scales polynomially with the value $\frac{1}{r_{\min} - 1}$.

In computational complexity theory, an algorithm whose runtime depends on the numeric value of the input (e.g., the magnitude of $1/(r-1)$) rather than just its bit length is classified as a \textbf{Pseudo-Polynomial Time Approximation Scheme (PPTAS)}. Therefore, even without the boundedness assumption, our algorithm remains efficient in the pseudo-polynomial sense, which is a standard and widely accepted efficiency benchmark for problems involving continuous parameters. The assumption is introduced primarily to claim the stronger FPTAS result presented in the main text.

\subsection{Grid Sufficiency (Discretization)}\label{f}
Since the aggregate production $A$ is a continuous variable, an exhaustive search is impossible. The following lemma validates our discretization strategy (Step 3 of Algorithm \ref{alg:FPTAS}). It ensures that the specific geometric grid $\mathcal{A}$ we constructed is dense enough to capture any potential equilibrium within the desired precision $\varepsilon$. The following is the proof of  Lemma \ref{select}

\begin{proof}
Because $\sigma_i$ and $A$ are one-to-one mappings on the interval $(v_k, v_{k+1}]$ for all $i \in I$, we rewrite $b_2(A, \sigma_i; t_i) = 0$ as $\sigma_i = \sigma(A, t_i)$ for notation convenience.

Let $\{A^*, I_A, \sigma^*\}$ be the $\varepsilon$-PNe, and let the two nodes next to $A^*$ be $A_1$ and $A_2$. Then $A_1 < A^* < A_2$.

By the definition of the $\varepsilon$-PNE,
\[
\sum_{i \in I_A} \sigma(A^*, t_i) \in \left(1-\varepsilon, 1 + \varepsilon\right).
\]

Because $\sigma(\cdot)$ is decreasing in $A$,
\[
\sum_{i \in I_A} \sigma(A_2, t_i) < \sum_{i \in I_A} \sigma(A^*, t_i) < \sum_{i \in I_A} \sigma(A_1, t_i).
\]

We show that either $A_1$ or $A_2$ constitutes an $\varepsilon$-PNE by contradiction. If neither $A_1$ nor $A_2$ constitutes an $\varepsilon$-PNE, then
\[
\sum_{i \in \mathcal{I}^A} \sigma(A_2, t_i) \leq 1 - \varepsilon, \quad \sum_{i \in \mathcal{I}^A} \sigma(A_1, t_i) \geq 1 + \varepsilon,
\]
which implies
\[
\sum_{i \in \mathcal{I}^A} \sigma(A_1, t_i) - \sum_{i \in \mathcal{I}^A} \sigma(A_2, t_i) \geq 2\varepsilon.
\]

On the other hand,
\begin{align*}
   &\sum_{i \in \mathcal{I}^A} \sigma(A_1, t_i) - \sum_{i \in \mathcal{I}^A} \sigma(A_2, t_i) \\
   &\leq \sum_{i \in \mathcal{I}^A} \max_{A\in[A_1,A_2]} \left\{ \left| \frac{\partial k_1(A; a_i)}{\partial A} \right|, \left| \frac{\partial k_2(A; a_i)}{\partial A} \right| \right\} \cdot (A_2-A_1)\\
   &\leq \sum_{i \in \mathcal{I}^A} \max_{A\in[A_1,A_2]}\left\{\frac{1}{\underline{r}'(\underline{r}'-1)}\frac{1}{A_1},\frac{1}{\underline{r}^2} \frac{1}{A_1}\right\}\cdot(A_2-A_1)\\
   & \leq \sum_{i \in \mathcal{I}^A} \max_{A\in[A_1,A_2]}\left\{\frac{1}{\underline{r}'(\underline{r}'-1)}\frac{1}{A_1},\frac{1}{\underline{r}^2} \frac{1}{A_1}\right\}\cdot\left(A_1 \cdot \min\{\underline{r}^2,\underline{r}'(\underline{r}'-1)\} \cdot \frac{\varepsilon}{n}\right)\\
   & \leq \varepsilon,
\end{align*}
which draws a contradiction.

The second inequality is shown in the proof of Lemma \ref{Lip} in Appendix \ref{g}.
\end{proof}

\subsection{Proof of Theorem \ref{npc2}}\label{E}

\begin{proof}
Membership in NP is immediate since any candidate strategy profile can be verified in polynomial time. In the following we show the hardness proof.

\medskip\noindent
\textbf{NP-hardness via padding.}
Let $\mathcal{G}$ be an arbitrary “all-medium” Tullock instance on \(m\) contestants (each \(r_i\in(1,2]\)), whose the existence of PNE is NP-complete by Lemma~\ref{npc}.  We build a new instance $\mathcal{G}'$ on
\[
n \;=\; m + m^d
\]
contestants by adding \(m^d\) dummy contestants, each with \(r_j\notin(1,2]\) and productivity \(a_j=\varepsilon\ll1\).  

Let 
\[
A_{\min}
\;=\;
\min\{A^* : A^* \text{ is the total effort in some equilibrium of the original }\mathcal{G}\}
\]
which is strictly positive.  Fix any dummy \(j\) with elasticity \(r_j\notin(1,2]\) and productivity \(a_j=\varepsilon\).  For any aggregate \(A\ge A_{\min}\), its payoff is
\[
u_j(x)
=\frac{\varepsilon\,x^{r_j}}{A + \varepsilon\,x^{r_j}}\,-\;x.
\]
\paragraph{Asymptotic bound for \(\max_x u_j(x)\).}  
Set \(r = r_j>1\), \(a = \varepsilon/A\).  For small \(\varepsilon\), the denominator \(A+\varepsilon x^r\approx A\), so
\[
u_j(x) = \frac{\varepsilon\,x^r}{A + \varepsilon x^r} - x
\;\approx\; a\,x^r - x.
\]
Let
\[
f(x) = a\,x^r - x.
\]
Then
\[
f'(x) = a\,r\,x^{r-1} - 1,
\]
and the unique interior critical point \(x^*>0\) satisfies
\[
a\,r\,(x^*)^{r-1} = 1
\quad\Longrightarrow\quad
x^* = \bigl(a\,r\bigr)^{-1/(r-1)}.
\]
Substituting back,
\[
f(x^*)
= a\,(x^*)^r - x^*
= a\,(x^*)^{r-1}\,x^* - x^*
= \frac{x^*}{r} \;-\; x^*
= -\frac{r-1}{r}\,x^*.
\]
Hence
\[
\max_{x\ge0}u_j(x)
\;\approx\;
f(x^*)
= -\frac{r-1}{r}\,\bigl(a\,r\bigr)^{-1/(r-1)}
= -\Theta\!\bigl(a^{-1/(r-1)}\bigr).
\]
Recalling \(a=\varepsilon/A=O(\varepsilon)\), we get
\[
\max_x u_j(x)
= -\Theta\!\bigl(\varepsilon^{-1/(r-1)}\bigr)
= O\!\bigl(\varepsilon^{-1/(r-1)}\bigr).
\]
Restoring the exact denominator only changes constants, so the same bound holds:
\[
\max_{x\ge0} u_j(x)
= O\!\bigl(\varepsilon^{-\tfrac{1}{r_j-1}}\bigr).
\]
In particular, as \(\varepsilon\to0\), \(\max_x u_j(x)\to0\), ensuring for small enough \(\varepsilon\) the dummy’s utility is always negative except at \(x=0\), so its unique best response is \(x_j=0\).

So
\[
\mathcal{G}\text{ has a PNE}
\;\Longleftrightarrow\;
\mathcal{G}'\text{ has a PNE}.
\]

\medskip\noindent
\textbf{Bounding \(m\) versus \(\log_2 n\).}  Since
\[
n = m + m^d \le 2\,m^d,
\]
we take base-2 logarithms:
\[
\log_2 n \;\le\;\log_2(2\,m^d)
= 1 + d\,\log_2 m
\quad\Longrightarrow\quad
\log_2 m \;\ge\;\frac{\log_2 n - 1}{d}.
\]
Exponentiating yields
\[
m \;\ge\; 2^{(\log_2 n - 1)/d}
= 2^{-1/d}\,n^{1/d}.
\]
Because
\(\lim_{n\to\infty} n^{1/d}/\log_2 n = \infty\), there exists \(N_0\) so that for all \(n\ge N_0\),
\[
2^{-1/d}\,n^{1/d} \;\ge\; \log_2 n.
\]
Hence for \(n\ge N_0\),
\[
m \;\ge\; \log_2 n,
\]
and in particular \(m \ge c\,\log_2 n\) for any fixed \(c<1\).

\medskip
Since $\mathcal{G}'$ can be constructed in polynomial time and the existence of PNE in $\mathcal{G'}$ is equivalent to that in the NP-complete instance $\mathcal{G}$, it follows that for all \(n\ge N_0\) with \(m\ge c\log_2 n\), deciding the existence of PNE is NP-complete.
\end{proof}

\subsection{Proof of Lemma \ref{select}}\label{f}
\begin{proof}
Because $\sigma_i$ and $A$ are one-to-one mappings on the interval $(v_k, v_{k+1}]$ for all $i \in I$, we rewrite $b_2(A, \sigma_i; t_i) = 0$ as $\sigma_i = \sigma(A, t_i)$ for notation convenience.

Let $\{A^*, I_A, \sigma^*\}$ be the $\varepsilon$-approximate solution, and let the two nodes next to $A^*$ be $A_1$ and $A_2$. Then $A_1 < A^* < A_2$.

By the definition of the $\varepsilon$-approximate solution,
\[
\sum_{i \in I_A} \sigma(A^*, t_i) \in \left(1-\varepsilon, 1 + \varepsilon\right).
\]

Because $\sigma(\cdot)$ is decreasing in $A$,
\[
\sum_{i \in I_A} \sigma(A_2, t_i) < \sum_{i \in I_A} \sigma(A^*, t_i) < \sum_{i \in I_A} \sigma(A_1, t_i).
\]

We show that either $A_1$ or $A_2$ constitutes an $\varepsilon$-approximate solution by contradiction. If neither $A_1$ nor $A_2$ constitutes an $\varepsilon$-approximate solution, then
\[
\sum_{i \in \mathcal{I}^A} \sigma(A_2, t_i) \leq 1 - \varepsilon, \quad \sum_{i \in \mathcal{I}^A} \sigma(A_1, t_i) \geq 1 + \varepsilon,
\]
which implies
\[
\sum_{i \in \mathcal{I}^A} \sigma(A_1, t_i) - \sum_{i \in \mathcal{I}^A} \sigma(A_2, t_i) \geq 2\varepsilon.
\]

On the other hand,
\begin{align*}
   &\sum_{i \in \mathcal{I}^A} \sigma(A_1, t_i) - \sum_{i \in \mathcal{I}^A} \sigma(A_2, t_i) \\
   &\leq \sum_{i \in \mathcal{I}^A} \max_{A\in[A_1,A_2]} \left\{ \left| \frac{\partial k_1(A; a_i)}{\partial A} \right|, \left| \frac{\partial k_2(A; a_i)}{\partial A} \right| \right\} \cdot (A_2-A_1)\\
   &\leq \sum_{i \in \mathcal{I}^A} \max_{A\in[A_1,A_2]}\left\{\frac{1}{\underline{r}'(\underline{r}'-1)}\frac{1}{A_1},\frac{1}{\underline{r}^2} \frac{1}{A_1}\right\}\cdot(A_2-A_1)\\
   & \leq \sum_{i \in \mathcal{I}^A} \max_{A\in[A_1,A_2]}\left\{\frac{1}{\underline{r}'(\underline{r}'-1)}\frac{1}{A_1},\frac{1}{\underline{r}^2} \frac{1}{A_1}\right\}\cdot\left(A_1 \cdot \min\{\underline{r}^2,\underline{r}'(\underline{r}'-1)\} \cdot \frac{\varepsilon}{n}\right)\\
   & \leq \varepsilon,
\end{align*}
which draws a contradiction.

The second inequality is shown in the proof of Lemma \ref{Lip} in Appendix \ref{g}.
\end{proof}

\subsubsection{Lipschitz Continuity of Utilities}\label{g}
Algorithm \ref{alg:FPTAS} identifies a state where the sum of action shares is close to 1 (Market Clearing). However, the definition of an $\varepsilon$-PNE requires that no player can gain more than $O(\varepsilon)$ utility by deviating. The following lemma bridges this gap. By establishing the Lipschitz continuity of the best-response and utility functions, we guarantee that a small error in the aggregate production $A$ translates to a bounded error in individual utilities.

\begin{lemma}\label{Lip}
    Let $ \underline{A} \triangleq \min_{i \in \mathcal{I}^2} \underline{A_i} $ and $ \bar{A} \triangleq \max_{i \in \mathcal{I}^2} \bar{A}_i $. The following continuity properties hold:
    \begin{enumerate}
        \item For a contestant $ i $, the best response action share $\sigma_i$ is Lipschitz continuous with respect to the aggregate production $A$. Specifically:
        \begin{itemize}
            \item If $ i \in \mathcal{I}^2 $, $\sigma_i$ is Lipschitz continuous on the interval $ A \in [\underline{A}, \bar{A}_i) \cup (\bar{A}_i, \bar{A}] $.
            \item If $ i \in \mathcal{I}^1 $, $\sigma_i$ is Lipschitz continuous on the interval $ A \in [\underline{A}, \bar{A}] $.
        \end{itemize}
        
        \item The utility function is Lipschitz continuous with respect to the aggregate production $A$. Specifically:
        \begin{itemize}
            \item If $ i \in \mathcal{I}^2 $, the utility function is Lipschitz continuous on the interval $ A \in [\underline{A}, \bar{A}_i) \cup (\bar{A}_i, \bar{A}] $.
            \item If $ i \in \mathcal{I}^1 $, the utility function is Lipschitz continuous on the interval $ A \in [\underline{A}, \bar{A}] $.
        \end{itemize}
    \end{enumerate}
\end{lemma}

\label{app:lip-proof}
\begin{proof}
We prove that the best-response share $\sigma_i$ and utility $u_i$ are Lipschitz continuous in aggregate production $A$, with precise constants and complete derivations.

\vspace{1ex}
\subsection*{(i) Lipschitz continuity of $\sigma_i$ with respect to $A$}

\paragraph{Case 1: $i \in \mathcal{I}^2$ (i.e., $r_i > 1$).}

In this case, the action share $\sigma_i$ is defined implicitly via the equation
\[
k_2(A, \sigma_i) := a_i r_i^{r_i} (1 - \sigma_i)^{r_i} \sigma_i^{r_i - 1} - A = 0,
\quad \text{for } A \in [\underline{A}, \bar{A}_i),
\]
and $\sigma_i(A) \equiv 0$ for $A \ge \bar{A}_i$.

We compute the derivative using implicit differentiation:
\[
\frac{d \sigma_i}{d A} = - \frac{\partial k_2 / \partial A}{\partial k_2 / \partial \sigma_i} = \frac{1}{\partial k_2 / \partial \sigma_i}.
\]

To simplify notation, let us denote:
\[
C := a_i r_i^{r_i}.
\]

From the equation we have:
\[
a_i r_i^{r_i}  = \frac{A}{(1 - \sigma_i)^{r_i} \sigma_i^{r_i - 1}}
\]

Then,
\[
\frac{\partial k_2}{\partial \sigma_i}
= C \left[ -r_i (1 - \sigma_i)^{r_i - 1} \sigma_i^{r_i - 1}
+ (r_i - 1)(1 - \sigma_i)^{r_i} \sigma_i^{r_i - 2} \right].
\]

Therefore, the derivative becomes:
\[
\frac{d \sigma_i}{d A}
= \frac{1}{C \left[ -r_i (1 - \sigma_i)^{r_i - 1} \sigma_i^{r_i - 1}
+ (r_i - 1)(1 - \sigma_i)^{r_i} \sigma_i^{r_i - 2} \right]}.
\]

We can further simplify this using algebra:
\[
\left| \frac{d \sigma_i}{d A} \right|
= \frac{1}{A \left( \frac{r_i}{1 - \sigma_i} - \frac{r_i - 1}{\sigma_i} \right)}.
\]

From known bounds:
\[
\sigma_i \in \left[ \frac{r_i - 1}{r_i}, 1 \right]
\Rightarrow 
\left( \frac{r_i}{1 - \sigma_i} - \frac{r_i - 1}{\sigma_i} \right) 
\in \left[r_i(r_i-1), \infty \right).
\]

Hence:
\[
\left| \frac{d \sigma_i}{d A} \right|
\le \frac{1}{r_i (r_i - 1)} \cdot \frac{1}{A}.
\]

Let $\underline{r}' = \min_{i \in \mathcal{I}^2} r_i > 1$. Then for all $i \in \mathcal{I}^2$:
\[
\left| \frac{d \sigma_i}{d A} \right| \le \frac{1}{\underline{r}' (\underline{r}' - 1) \cdot \underline{A}} =: L_1.
\]

Thus, for all $A_1, A_2 \in [\underline{A}, \bar{A}_i)$,
\[
|\sigma_i(A_1) - \sigma_i(A_2)| \le L_1 |A_1 - A_2|.
\]

\paragraph{Case 2: $i \in \mathcal{I}^1$ (i.e., $r_i < 1$).}

The best–response share $\sigma_i$ is defined implicitly by the first–order condition:
\[
  k_1(\sigma, A) := (1 - \sigma)  - \frac{A}{r_i a_i} \left( \frac{\sigma A}{a_i} \right)^{\alpha} = 0,
  \quad\text{where } \alpha := \tfrac{1}{r_i} - 1 > 0.
\]

Implicit differentiation gives
\[
  \left| \frac{d\sigma_i}{dA} \right|
  = -\frac{\partial k_1 / \partial A}{\partial k_1 / \partial \sigma}
  = \frac{
      \tfrac{1}{r_i a_i} \left( \tfrac{\sigma A}{a_i} \right)^{\alpha}
    + \tfrac{\alpha \sigma A}{r_i a_i^2} \left( \tfrac{\sigma A}{a_i} \right)^{\alpha - 1}
  }{
      1 + \tfrac{\alpha A^2}{r_i a_i^2} \left( \tfrac{\sigma A}{a_i} \right)^{\alpha - 1}
  }. \tag{S.0}
\]

We simplify this expression step-by-step.

\begin{align*}
\left| \frac{d\sigma_i}{dA} \right|
&= \frac{\displaystyle
        \frac{1}{r_i a_i} \left( \tfrac{\sigma A}{a_i} \right)^{\alpha}
      + \frac{\alpha \sigma A}{r_i a_i^2} \left( \tfrac{\sigma A}{a_i} \right)^{\alpha - 1}
     }{\displaystyle
        1 + \frac{\alpha A^2}{r_i a_i^2} \left( \tfrac{\sigma A}{a_i} \right)^{\alpha - 1}
     }
\\[6pt]
&= \frac{\displaystyle
        \frac{\sigma A}{r_i^2 a_i^2} \left( \tfrac{\sigma A}{a_i} \right)^{\alpha - 1}
     }{\displaystyle
        1 + \frac{\alpha A^2}{r_i a_i^2} \left( \tfrac{\sigma A}{a_i} \right)^{\alpha - 1}
     }
\\[6pt]
&= \frac{
        \sigma^{\alpha} A^{\alpha}
     }{
        r_i^2 a_i^{\alpha + 1} + \alpha r_i \sigma^{\alpha - 1} A^{\alpha + 1}
     }.
\end{align*}

Using $0 < \sigma \le 1$, we can bound
\[
  \left| \frac{d\sigma_i}{dA} \right|
  \le \frac{\sigma^{\alpha} A^{\alpha}}{ r_i^2 a_i^{\alpha + 1}}
  \le \frac{A^{\alpha}}{ r_i^2 a_i^{\alpha + 1}}.
\]

Letting $\underline{r} = \min_{i \in \mathcal{I}^1} r_i$ and $\underline{a} = \min_i a_i$, and using $A \le \bar{A}$, we obtain the final bound:
\[
  \left| \frac{d\sigma_i}{dA} \right|
  \le \frac{\bar{A}^{\frac{1}{\underline{r}} - 1}}{ \underline{r}^2 \, \underline{a}^{\frac{1}{\underline{r}}}} =: L_2.
\]

This completes the proof that $\sigma_i$ is Lipschitz continuous in $A$ for $r_i < 1$ with constant $L_2$.

\paragraph{Final Bound.}
Let
\[
L_\sigma := \max(L_1, L_2) = \max \left(\frac{1}{\underline{r}' (\underline{r}' - 1) \cdot \underline{A}},\frac{\bar{A}^{\frac{1}{\underline{r}} - 1}}{ \underline{r}^2 \, \underline{a}^{\frac{1}{\underline{r}}}} \right)
\]
denote the global Lipschitz constant for $\sigma_i$ with respect to $A$, covering both $\mathcal{I}^1$ and $\mathcal{I}^2$ cases.

\subsection*{(ii) Lipschitz continuity of the utility function with respect to $A$.} 

Each contestant's utility is given by
\[
u_i(A) = \sigma_i(A) - x_i(A),
\qquad \text{where } x_i(A) = \left( \frac{\sigma_i(A)\,A}{a_i} \right)^{1/r_i}.
\]

We analyze the variation $|u_i(\hat{A}) - u_i(A^*)|$ under the assumption that $|\sigma_i(\hat{A}) - \sigma_i(A^*)| \le \varepsilon$ and $|\hat{A} - A^*| \le \varepsilon$.

First term: the difference in reward shares is bounded by
\[
\left| \sigma_i(\hat{A}) - \sigma_i(A^*) \right| = |\sigma_i(\hat{A}) - \sigma_i(A^*)| \le L_{\sigma}\,\varepsilon.
\]

Second term: define
\[
X := \frac{\sigma_i(A^*)\,A^*}{a_i},
\qquad
\hat{X} := \frac{\sigma_i(\hat{A})\,\hat{A}}{a_i}.
\]

Then
\[
|x_i(\hat{A}) - x_i(A^*)|
= \left| \hat{X}^{1/r_i} - X^{1/r_i} \right|.
\]

Since $t \mapsto t^{1/r_i}$ is differentiable and increasing for $t > 0$, and both $X, \hat{X}$ lie in the bounded interval
\[
\left[\frac{\underline{\sigma}\,\underline{A}}{\bar{a}}, \, \frac{\bar{\sigma}\,\bar{A}}{\underline{a}} \right]
\subseteq \left[0, \, \frac{\bar{a}\,\bar{A}}{\underline{a}} \right],
\]
we use the Lipschitz continuity of the map $t^{1/r_i}$ on this interval. Its Lipschitz constant is
\[
L' := \max_{i} \frac{1}{r_i} \left( \frac{\bar{a}\,\bar{A}}{\underline{a}} \right)^{\frac{1 - r_i}{r_i}} 
\le \frac{1}{\underline{r}} \left( \frac{\bar{a}\,\bar{A}}{\underline{a}} \right)^{\frac{1 - \underline{r}}{\underline{r}}}.
\]

Next, we bound $|\hat{X} - X|$:
\[
|\hat{X} - X|
= \left| \frac{\sigma_i(\hat{A})\,\hat{A} - \sigma_i(A^*)\,A^*}{a_i} \right|
\le \frac{1}{\underline{a}} \left( |\hat{A} - A^*| + \bar{A}\,|\sigma_i(\hat{A}) - \sigma_i(A^*)| \right)
\le \frac{L_{\sigma}\bar{A}+1}{\underline{a}}\,\varepsilon =: L''\,\delta.
\]

Thus the second term is bounded as
\[
|x_i(\hat{A}) - x_i(A^*)| \le L' L'' \varepsilon.
\]

\vspace{0.5ex}
\paragraph{Final Bound.} Combining both parts,
\begin{align}\label{lipschize}
|u_i(\hat{A}) - u_i(A^*)|
& \le \varepsilon + L' L'' \varepsilon = \left( 1 + L' L'' \right) \varepsilon \\
& = \left(1+ \frac{1}{\underline{a}\,\underline{r}} \left( \frac{\bar{a}\,\bar{A}}{\underline{a}} \right)^{\frac{1 - \underline{r}}{\underline{r}}}\max \left(\frac{1}{\underline{r}' (\underline{r}' - 1) \cdot \underline{A}},\frac{\bar{A}^{\frac{1}{\underline{r}} - 1}}{ \underline{r}^2 \, \underline{a}^{\frac{1}{\underline{r}}}} \right) \bar{A} + 1\right) \varepsilon\\
&=L_u \varepsilon.
\end{align}

\end{proof}

\subsubsection{Proof of Main FPTAS Theorem}\label{gg}
Finally, we synthesize the results above to prove the main algorithmic theorem. The proof combines the complexity analysis (showing the grid size and subset sum routine are polynomial) with the correctness analysis (using Lemma \ref{select} and Lemma \ref{Lip} to establish the $(L\varepsilon)$-PNE property). The following is the proof of Theorem \ref{thm:fptas_main}

\begin{proof}
We prove the first part of Theorem \ref{thm:fptas_main} by constructing a polynomial-time algorithm (Algorithm \ref{alg:FPTAS}) to compute an $\varepsilon$-approximate solution. The algorithm systematically partitions the range of aggregate actions $ A $ and verifies potential solutions within each interval. Let $ \underline{A} $ and $ \bar{A} $ be as defined earlier in Eq. \eqref{eq:thresholds}. The algorithm examines the three distinct intervals of $A$:

\paragraph{Interval $(0, \underline{A}]$}: In this interval, all contestants, including those in $\mathcal{I}^2$, participate. Since the active contestant set is fixed, a PNE can be efficiently identified using binary search, which also guarantees an $\varepsilon$-approximate solution with polynomial time.

\paragraph{Interval $(\bar{A}, A_{\max})$}: In this interval, no contestant in $\mathcal{I}^2$ participates. The active contestant set in $\mathcal{I}^1$ is determinable, and binary search can again be employed to find a PNE, ensuring an $\varepsilon$-approximate solution with polynomial time.

\paragraph{Interval $(\underline{A}, \bar{A}]$}: To address this, a set of candidate nodes is constructed for verification. This set, denoted as $\mathcal{A} = \text{sorted}(A_1 \cup A_2)$. The total number of nodes in $ \mathcal{A} $ is bounded by $ |\mathcal{A}| = 2n_2 + \log_{1+ \min\{\underline{r}^2,\underline{r}'(\underline{r}'-1)\} \cdot \frac{\varepsilon}{n}} \frac{\bar{A}}{\underline{A}}$, which is $O(\frac{n}{\varepsilon})$.

By Lemma \ref{select}, if there exists an $ \varepsilon $-approximate solution in the interval, then at least one of its neighboring nodes constitutes an $ \varepsilon $-approximate solution. Therefore it is sufficient to only verify the selected nodes to determine the existence of $ \varepsilon $-approximate solution.

Lastly, we explain the process of verifying an $ \varepsilon $-approximate solution at each selected node $ A_i $ in polynomial time. For illustration, consider a node $ A_i \in (\underline{A}_2, \bar{A}_1] $. We begin by computing the action shares of the active contestants from contestants in $ \mathcal{I}^1 $, as well as the action shares of active contestants from $ \mathcal{I}^2 $. These active contestants' action shares are denoted collectively as the set $ S_0 $.

Next, we analyze the remaining contestants in $ \mathcal{I}^2 $. For these contestants, we can exclude those who are guaranteed not to be active contestants based on their properties. For the remaining $ k $ contestants in $ \mathcal{I}^2 $, each can either participate ($ \sigma_i > 0 $) or not ($ \sigma_i = 0 $), leading to $ 2^k $ possible combinations of action shares. Since this exponential growth is computationally infeasible for large $ k $, a direct enumeration of all combinations is impractical.

To compute an $\varepsilon$-approximate solution in polynomial time, we consider the scenario where all $k$ remaining contestants participate. Specifically, let $ S = \{\sigma_i, \dots, \sigma_j\} $ denote the set of their action shares, where each share satisfies the best-response condition $ b_2(A, \sigma_i; a_i, r_i) = 0, \dots, b_2(A, \sigma_j; a_j, r_j) = 0 $. 

Since the best-response action share does not admit a closed-form expression when
$r_i\neq 1$, each $\sigma_i$ must be computed using a numerical root-finding routine.
This routine is a one-dimensional refinement procedure
that returns a $\delta$-accurate solution in
$O\!\left(\log\frac{1}{\delta}\right)$ iterations.

To ensure that the deviation in the aggregate constraint
$\sum_i \sigma_i = 1$ is within $\varepsilon$, we note that errors in individual action
shares accumulate linearly across contestants. Hence, it suffices to compute each
$\sigma_i$ to precision $\delta = \Theta(\varepsilon/n)$, which guarantees that
$\bigl|\sum_i \sigma_i - 1\bigr| \le \varepsilon$.

With this choice of $\delta$, the cost of computing a single best-response action
share $\sigma_i$ is
$O\!\left(\log\frac{n}{\varepsilon}\right)$.
Therefore, computing the action shares of all $n$ contestants at a fixed aggregate
production level $A$ requires
$O\!\left(n \log\frac{n}{\varepsilon}\right)$ time.
The problem then reduces to determining whether there exists a subset of $S$ whose
total sum, together with $S_0$, lies in the interval $(1-\varepsilon,1+\varepsilon)$.

To solve this subset selection problem, we use the \textit{Approximate Subset Sum Algorithm}, a well-established algorithm in the literature (e.g., \cite{Mart1990}). However, most existing subset sum algorithms are designed for integer elements, whereas our problem involves continuous values. To address this, we employ the \textit{Merge-and-Trim Approximation}, as described in \cite{KellererPferschyPisinger2004}, which efficiently handles continuous elements while maintaining computational feasibility. This method solves the problem in time $ O(\frac{n^3}{\varepsilon} \log^2 n) $. The key idea is to dynamically construct multiple subsets while systematically discarding intermediate subsets whose current sums are very close to one another. This pruning strategy ensures that the algorithm remains efficient by keeping the number of subsets manageable.

By applying this algorithm, we efficiently verify whether a given node $ A_i $ admits an $ \varepsilon $-approximate solution, thereby making the computation of approximate solutions feasible within polynomial time. This method leverages the special structure of the Tullock contest and provides a practical approach for handling otherwise computationally intractable scenarios.

Algorithm \ref{subset} describes the verification of $ \varepsilon $-approximate solution. One thing to notice is, at a given aggregated action $ A $, there can be active contestants, non-active contestants, and uncertain contestants. So that the initial subset should include all participating contestants, and the selection happens among uncertain contestants. 

After obtaining a set of $\varepsilon$-approximate solutions, we further verify whether each candidate satisfies the conditions of an $\varepsilon$-PNE. Specifically, for each agent, we fix the efforts of all other agents and solve for its best response. Since this is a single-variable optimization problem, finding the optimal best effort is trivial. If all agents' current efforts are within $\varepsilon$ of their respective best responses, the solution is confirmed as a valid $\varepsilon$-PNE. Otherwise, it is discarded. This final verification step ensures that the returned equilibria are robust and truly satisfy the $\varepsilon$-PNE conditions.

In summary, if a PNE exists, our algorithm outputs at least one
$\varepsilon$-approximate solution. It suffices to verify a set of
$O\!\left(\frac{n}{\varepsilon}\right)$ selected nodes.
For each node, computing the best-response action shares of all contestants
requires
$O\!\left(n\log\frac{n}{\varepsilon}\right)$ time, while the subsequent
subset sum verification using the merge-and-trim approximation algorithm
runs in
$O\!\left(\frac{n^3}{\varepsilon}\log^2 n\right)$ time.
Therefore, the overall running time of the algorithm is
\[
O\!\left(\frac{n^4}{\varepsilon^2}\log^2 n\right),
\]
which is polynomial in both $n$ and $1/\varepsilon$.

We prove the second part of Theorem \ref{thm:fptas_main} by analyzing the aggregate action $ A^* $ corresponding to a PNE and demonstrating that the $\varepsilon$-solution identified by Algorithm~\ref{alg:FPTAS} constitutes an $(L\varepsilon)$-Nash equilibrium.

\paragraph{Case 1: $ A^* < \underline{A} $ or $ A^* > \bar{A} $}  
In this scenario, Algorithm~\ref{alg:FPTAS} directly computes the PNE. By definition, an exact PNE trivially satisfies the conditions of an $(L\varepsilon)$-Nash equilibrium, thus establishing the proposition in this case.
\paragraph{Case 2: $ A^* \in [\underline{A}, \bar{A}] $}  
When $ A^* $ lies within the range $[\underline{A}, \bar{A}]$, we split the proof into two subcases.
\\ \textbf{Subcase 2.1: $ A^* $ is on the boundary of some contestant $ i \in \mathcal{I}^2 $ ($ A^* = \underline{A_i} $ or $ A^* = \bar{A_i} $)}. 
Since Algorithm~\ref{alg:FPTAS} enumerates all the boundaries of contestant $i \in \mathcal{I}^2$, any $ A^* $ that falls exactly on one of these boundaries will be directly found as an output by the algorithm. In this case, the equilibrium is trivially an $L\varepsilon$-Nash equilibrium.
\\
\textbf{Subcase 2.2: $ A^* $ is not on the boundary of any contestant $ i \in \mathcal{I}^2 $ ($ A^* \neq \underline{A_i}, A^* \neq \bar{A_i} $)}. 
In this scenario, we aim to show that there exists an $ A' $ close to $ A^* $, where $ A' $ serves as an $\varepsilon$-approximate solution. By Lemma~\ref{select}, $ A' $ will be covered within the set of candidate nodes examined by Algorithm~\ref{alg:FPTAS}. Furthermore, according to Lemma~\ref{Lip}, the action share for each contestant is Lipschitz continuous with respect to $ A $, and we know that the action share function of the active contestants is monotonic. Therefore, there exists an $ A' \in (A^* - \frac{\varepsilon}{L_{\sigma}n}, A^* + \frac{\varepsilon}{L_{\sigma}n}) $ such that the active contestants under $ A' $ are the same as those under $ A^* $, and the sum of action shares satisfies $ \sum_i \sigma_{i,A'} \in (1 - \varepsilon, 1 + \varepsilon) $.

By Lemma~\ref{Lip}, the utility function is Lipschitz continuous with respect to $ A $, and thus the deviation in utility function between $ A^* $ and $ A' $ is bounded by:
\[
u_{i}(A') - u_{i}(A^*) < L_u |A' - A^*| < L_u \frac{\varepsilon}{L_{\sigma}n} = L\varepsilon.
\]

In such case, the approximate solution $ A' $ found by the algorithm is an $(L\varepsilon)$-Nash equilibrium.
\\
Hence, we have shown that if a PNE exists, at least one of the $\varepsilon$-approximate solutions identified through Algorithm~\ref{alg:FPTAS} constitutes an $(L\varepsilon)$-Nash equilibrium. 

\end{proof}

\renewcommand{\thealgocf}{\arabic{algocf}} 
\setcounter{algocf}{0}
\addtocounter{algocf}{1}
\section{Algorithmic Descriptions}\label{ALG}
\SetAlgoNoEnd

This section details the specific algorithmic procedures referenced in the main text. We present the pseudocode for the exact solver in the logarithmically bounded regime, followed by the approximation subroutines used in our FPTAS for the general intractable regime.

\subsection{Unified Exact Algorithm for Limited Heterogeneity}

Algorithm \ref{alg:pne_tullock} formalizes the polynomial-time search procedure described in Section 4.3. This algorithm is designed for the regime where the number of medium-elasticity contestants is small ($m = O(\log n)$).

\begin{algorithm}[H]\label{polyalg}
\SetAlgoNoEnd
\caption{Identifying PNE in Tullock Contests with at Most $ m $ Medium Elasticity contestants}
\label{alg:pne_tullock}
\KwIn{Number of contestants $ n $, a vector of efficiency $ \bm{a}$, a vector of elasticity $ \bm{r} $, maximum medium-elasticity count $ m $, approximation parameter $ \varepsilon $}
\KwOut{A PNE list $ Y $ if it exists, or a statement that no PNE exists.}

Compute $ \underline{A_i} $ and $ \bar{A_i} $ for all contestants with $ r_i > 1 $\;
\ForEach{contestant $ j $ with $ r_j > 2 $}{ 
\tcp{Assume contestant $ j $ is active}
    $\mathcal{I}' \gets \{ i \mid r_i \in (1,2] \}$\;
    \tcp{Enumerate all possible combinations of active sets for medium contestants}
    $\mathcal{S} \gets \mathcal{P}(\mathcal{I}') = \{ S_k \mid S_k \subseteq \mathcal{I}', \, k = 1, \dots, 2^m \}$\;
    \ForEach{node $S \in \mathcal{S}$}{
        $\mathcal{I}^A \gets \mathcal{I}^1 \cup S \cup \{j\}$, $\mathcal{I}^{-A} = \mathcal{I}-\mathcal{I}^A$ \tcp{Determine active and inactive contestants}
        $ A_1 \gets  \max(\underline{A_i}, i \in \mathcal{I}^A)$, $A_2 \gets \min(\bar{A_i},i \in \mathcal{I}^{-A})$ \tcp{Compute the valid interval}
        \If{$[A_1,A_2]$ is a valid interval}{
            Perform binary search for $ A^* $ within this interval\;
            \While{$ A^* $ not found}{
                Compute $ S(A^*) = \sum_{i \in \mathcal{I}^A} \sigma_i $\;
                \If{$ S(A^*) \in [1-\frac{\varepsilon}{nL},1+\frac{\varepsilon}{nL}] $}{
                    Append the PNE solution to $ Y $\;
                }
            }
        }
    }
}
\If{$ Y \neq \varnothing $}{
    \Return $ Y $ as the PNE solution\;
}
\Return No PNE exists\;
\end{algorithm}

\subsection{Approximation Subroutines for FPTAS}

In the general regime where $m$ is large, exact enumeration is impossible. As detailed in Section 5.3, our FPTAS relies on discretizing the aggregate production space and verifying market clearing at each grid point. This verification reduces to a variant of the Subset Sum problem over continuous values.

Algorithm \ref{subset} describes the \texttt{APPROX-SUBSET-SUM} procedure. It determines whether a subset of "uncertain" contestants (those with $r_i \in (1,2]$ whose participation status is ambiguous at the current $A$) can be combined with the "mandatory" contestants (set sum $S_0$) to yield a total action share within the target range $(1-\varepsilon, 1+\varepsilon)$.

\begin{algorithm}[H]
\SetAlgoNoEnd
\caption{APPROX-SUBSET-SUM($S_0, \bm{\sigma}, \varepsilon$)}
\label{subset}
\KwIn{$ S_0 $: The sum of the action shares of contestants who participate for sure;  
$\bm{\sigma}$: A vector of action shares of uncertain contestants if they participate;  
$\varepsilon$: Approximation parameter.}
\KwOut{If there exist subsets of $\bm{\sigma}$ that together with $S_0$ sum up to within $\left( 1-\varepsilon, 1+\varepsilon \right)$, return at least one of them; otherwise, return nothing.}

$ n \gets |\bm{\sigma}| $ \tcp{Number of uncertain contestants}
$ X_0 \gets [S_0] $, $ Y_0 \gets [S_0] $ \tcp{Initialize solution sets}

\For{$ i \gets 1 $ \textbf{to} $ n $}{
    $ X_i \gets \texttt{MergeList}(X_{i-1}, X_{i-1} + \sigma_i) $\;
    $ X_i \gets \texttt{TRIM-FROM-BELOW}(X_i, \frac{\varepsilon}{2n}) $\;
    $ Y_i \gets \texttt{MergeList}(Y_{i-1}, Y_{i-1} + \sigma_i) $\;
    $ Y_i \gets \texttt{TRIM-FROM-ABOVE}(Y_i, \frac{\varepsilon}{2n}) $\;
}

\Return all elements in $ X_n $ and $ Y_n $ that are in $ \left( 1-\varepsilon, 1+\varepsilon \right) $\;
\end{algorithm}

Unlike the standard dynamic programming approach for integer subset sum, Algorithm \ref{subset} handles continuous variables using a list-based \textbf{Merge-and-Trim} strategy. This approach maintains a sorted list of reachable partial sums. In each step , it generates a new list by adding the -th candidate share  to all existing sums, merges it with the previous list, and then "trims" the result to keep the list size polynomial.

\subsection{Trimming Operations}

The efficiency of the \texttt{APPROX-SUBSET-SUM} routine hinges on the trimming operations, which prune the list of partial sums to prevent exponential growth while preserving approximation guarantees. These procedures ensure that the lists act as a \textit{sparse representation} of the solution space.

\begin{algorithm}[H]
\SetAlgoNoEnd
\caption{TRIM-FROM-BELOW($L, \delta$)}
\label{alg:trim_below}
\KwIn{$ L $: a list of numbers, $\delta$: trim parameter.}
\KwOut{A trimmed list $ L' $ with redundant values removed.}

Sort $ L $ in ascending order\;
$ last \gets L[1] $ \tcp{Initialize reference value}
$ L' \gets [last] $ \tcp{Initialize trimmed list}

\For{$ i \gets 2 $ \textbf{to} $ |L| $}{
    \If{$ L[i] > (1 + \delta) \cdot last $}{
        Append $ L[i] $ to $ L' $\;
        $ last \gets L[i] $\;
    }
}
\Return $ L' $\;
\end{algorithm}

\begin{algorithm}[H]
\SetAlgoNoEnd
\caption{TRIM-FROM-ABOVE($L, \delta$)}
\label{alg:trim_above}
\KwIn{$ L $: a list of numbers, $\delta$: trim parameter.}
\KwOut{A trimmed list $ L' $ with redundant values removed.}

Sort $ L $ in descending order\;
$ last \gets L[1] $ \tcp{Initialize reference value}
$ L' \gets [last] $ \tcp{Initialize trimmed list}

\For{$ i \gets 2 $ \textbf{to} $ |L| $}{
    \If{$ L[i] < last \cdot (1 - \delta) $}{
        Append $ L[i] $ to $ L' $\;
        $ last \gets L[i] $\;
    }
}
\Return $ L' $\;
\end{algorithm}

Algorithms \ref{alg:trim_below} and \ref{alg:trim_above} detail these specific implementations:
\begin{itemize}
    \item \textbf{\texttt{TRIM-FROM-BELOW} (Algorithm \ref{alg:trim_below}):} This procedure processes the list in ascending order to establish a lower-bound approximation. It initializes with the smallest element and iterates through the list, retaining a new value $L[i]$ only if it sufficiently exceeds the previously kept value $last$. Specifically, the condition for retention is:
    \[
    L[i] > last \cdot (1 + \delta).
    \]
    This ensures that any value discarded is approximated by a kept value within a factor of $(1+\delta)$, limiting the list size to be logarithmic in the value range.

    \item \textbf{\texttt{TRIM-FROM-ABOVE} (Algorithm \ref{alg:trim_above}):} This procedure processes the list in descending order. It initializes with the largest element and retains a subsequent smaller value $L[i]$ only if it drops sufficiently below the current reference $last$. The retention condition is:
    \[
    L[i] < last \cdot (1 - \delta).
    \]
    This guarantees geometric sparsification from the upper end, ensuring efficient handling of the solution space's upper bounds.
\end{itemize}

By applying these trimming operations at every step of the subset sum construction (Algorithm \ref{subset}), we ensure that the number of candidate sums remains polynomial in $n$ and $1/\varepsilon$, satisfying the runtime requirements of the FPTAS.

\end{document}